\documentclass[trackchanges,twocolumn]{aastex701}
\usepackage{verbatim}
\usepackage[T1]{fontenc}
\usepackage[utf8]{inputenc}
\usepackage[american]{babel}
\usepackage{epsfig}
\usepackage{booktabs}
\usepackage{multirow}
\usepackage{dcolumn}
\usepackage{amsmath}
\usepackage{mathtools}
\usepackage{amsfonts}
\usepackage{amssymb}
\usepackage{ulem}
\usepackage{epstopdf}
\usepackage{bm}
\usepackage{siunitx}
\usepackage{braket}
\usepackage{enumitem}
\usepackage{soul}
\usepackage[table]{xcolor}
\usepackage{color}
\usepackage{transparent}
\usepackage{pifont}
\usepackage{enumitem}
\usepackage{float}

\definecolor{navyblue}{rgb}{0.0, 0.0, 0.5}
\definecolor{royalblue}{rgb}{0.25, 0.41, 0.88}
\definecolor{cadmiumgreen}{rgb}{0.0, 0.42, 0.24}
\definecolor{blue-violet}{rgb}{0.54, 0.17, 0.89}
\definecolor{darkviolet}{rgb}{0.58, 0.0, 0.83}
\definecolor{orange(colorwheel)}{rgb}{1.0, 0.5, 0.0}

\usepackage{hyperref}

\usepackage{booktabs}
\usepackage{multirow}
\usepackage{dcolumn}
\usepackage{colortbl}
\usepackage{orcidlink}

\begin{document}

\title{Cosmographic Footprints of Dynamical Dark Energy}

\author[orcid=0009-0005-0594-4128,sname='Fazzari']{Elisa Fazzari}
\affiliation{Physics Department, Sapienza University of Rome, P.le A. Moro 5, 00185 Roma, Italy}
\affiliation{Istituto Nazionale di Fisica Nucleare (INFN), Sezione di Roma, P.le A. Moro 5, I-00185, Roma, Italy}
\affiliation{Physics Department, Tor Vergata University of Rome, Via della Ricerca Scientifica 1, 00133 Roma, Italy}
\email[show]{elisa.fazzari@uniroma1.it}  

\author[orcid=0000-0002-4012-9285,gname='William', sname='Giar\`e']{William Giar\`e} 
\affiliation{School of Mathematical and Physical Sciences, University of Sheffield, Hounsfield Road, Sheffield S3 7RH, United Kingdom}
\email[show]{w.giare@sheffield.ac.uk}

\author[orcid=0000-0001-8408-6961,gname='Eleonora', sname='Di Valentino']{Eleonora Di Valentino} 
\affiliation{School of Mathematical and Physical Sciences, University of Sheffield, Hounsfield Road, Sheffield S3 7RH, United Kingdom}
\email{e.divalentino@sheffield.ac.uk}

\begin{abstract}
We introduce a novel cosmographic framework to trace the late-time kinematics of the Universe without assuming any underlying dynamics. The method relies on generalized Padé-$(2,1)$ expansions around arbitrary pivot redshifts, which, compared to state-of-the-art calculations, reduce truncation errors by up to two orders of magnitude at high redshift and yield more precise constraints by defining cosmographic parameters exactly where the data lie. This avoids extrapolations, mitigates degeneracies, and enables a clean disentangling of their effects. Using the latest low-redshift datasets, we center the generalized expansion in multiple bins across $z\in[0,1]$ and obtain precise constraints on the redshift evolution of cosmographic parameters. We find that all key parameters deviate from their $\Lambda$CDM predictions in a redshift-dependent way that can be naturally explained within dynamical Dark Energy scenarios. The deceleration parameter $q(z)$ follows a redshift evolution consistent with the Chevallier–Polarski–Linder (CPL) parameterization, while the generalized $Om(z)$ diagnostic shows deviations of up to $\sim4\sigma$ from the constant $\Lambda$CDM expectation, closely matching the CPL predictions. Taken together, these results point to footprints of dynamical Dark Energy in the kinematics of the Universe at $z\lesssim 1$.
\end{abstract}

\section{Introduction}
Baryon Acoustic Oscillation (BAO) measurements released by the Dark Energy Spectroscopic Instrument~\citep{DESI:2024mwx,DESI:2024uvr,DESI:2024kob,DESI:2024lzq,DESI:2024aqx,DESI:2025zgx,DESI:2025zpo,DESI:2025qqy} (DESI) have revitalized interest in one of the most profound open problems in cosmology and fundamental physics: the nature of the current accelerated expansion of the Universe. 

DESI percent-level constraints on the transverse comoving distance, the Hubble rate, and their combination (all relative to the sound horizon at the drag epoch, $r_d$) directly challenge the cosmological constant ($\Lambda$) interpretation of cosmic acceleration within the standard $\Lambda$ Cold Dark Matter ($\Lambda$CDM) model, pointing to a preference for evolving Dark Energy (DE)~\citep{DESI:2025zgx,DESI:2025wyn}. 

This has sparked intense debate and a wide range of new analyses and interpretations that, broadly speaking, can be grouped into two main categories~\citep{Sahni:2006pa}. The first comprises model-dependent analyses, where one assumes a theoretical framework to describe the late-time dynamics of the Universe, for instance, by specifying how the DE equation of state (EoS), $w(a)$, evolves with expansion. This is the strategy first adopted by the DESI collaboration, which, under the assumption of a Chevallier–Polarski–Linder (CPL) parametrization, $w(a) = w_0 + w_a(1-a)$~\citep{Chevallier:2000qy,Linder:2002et}, reported a preference for evolving DE at the level of $\sim 2.8\sigma$–$4.2\sigma$, depending on the specific combination of DESI BAO, Type Ia supernovae (SNIa), and Cosmic Microwave Background (CMB) data~\citep{DESI:2025zgx,DESI:2025wyn}; see also~\citep{Giare:2024oil,Giare:2025pzu}. While this preference remains fairly robust across different parameterizations~\citep{Giare:2024gpk,Wolf:2025jlc}, more exotic scenarios (or the inclusion of additional free parameters) can shift the evidence from negligible~\citep{Nesseris:2025lke} to $\gtrsim 5\sigma$~\citep{Scherer:2025esj}, laying bare the inherent dependence of such conclusions on the assumed \textit{ansatz}.

A second assumption-free approach is to bypass any parametrization and reconstruct quantities such as $w(a)$ directly from data via non-parametric or Machine Learning techniques; see, e.g., ~\citet{Goh:2024exx,Ormondroyd:2025exu,Ormondroyd:2025iaf,Berti:2025phi,DESI:2025fii,Gonzalez-Fuentes:2025lei, Mukherjee:2025ytj}. These methods, however, face other well-known challenges: reconstructions tend to overfit noise, depend on binning choices, and generally suffer from reduced precision and large uncertainties, preventing them from delivering conclusive statements about the nature of DE.

Yet a third possible approach -- potentially able to capture the advantages of both model-dependent and model-independent analyses -- is provided by cosmography~\citep{Sahni:2002fz,Visser:2003vq,Visser:2004bf,Vitagliano:2009et,Bamba:2012cp,Aviles:2012ay,Dunsby:2015ers,Capozziello:2020ctn}. Unlike traditional methods, cosmography makes no assumptions about the underlying dynamics of the Universe~\citep{Visser:2004bf,Dunsby:2015ers}, and instead expands observable quantities such as the expansion rate $H(z)$ or the luminosity distance $D_{\rm L}(z)$, which encode its geometry and kinematics. These functions are typically expanded around $z_0 = 0$ (the present epoch), using a Taylor series or, preferably, a Padé approximation to improve convergence at higher redshift~\citep{Gruber:2013wua,Wei:2013jya,Aviles:2014rma}. The coefficients of the expansion map directly onto cosmographic parameters with clear physical interpretation, such as the present-day expansion rate $H_0$, the deceleration parameter $q_0$, and, at higher order, the jerk $j_0$ and the snap $s_0$. Once specified, these parameters uniquely fix the truncated series for $H(z)$, $D_{\rm L}(z)$, and related quantities at higher redshift, which can then be fitted to BAO, SNIa, and other low-redshift data, yielding constraints with a precision competitive with specific models.

While elegant, cosmography is not free of limitations~\citep{Busti:2015xqa}. In its standard form, it relies on truncated series expansions of kinematic and geometric quantities around $z_0 = 0$. The fitted parameters (e.g., $H_0$, $q_0$, $j_0$, $s_0$) describe the Universe at the present epoch, while the evolution at $z > 0$ is inferred indirectly from the truncated expansion. At higher redshift, the approximation deteriorates and truncation errors can become comparable to, or even larger than, observational uncertainties -- precisely where most of the data lie (i.e., $0.1 \lesssim z \lesssim 2$). A common workaround is to extend the expansion to higher orders, but this introduces additional free parameters, leading to degeneracies and reduced predictive power. In practice, this entails a double risk: either we fail to exploit the full potential of the data or, worse, introduce biases.

To overcome these limitations, we introduce a generalized cosmographic framework that allows us to trace the kinematics and expansion history of the Universe up to $z \gtrsim 1$ with unprecedented precision, revealing clear departures from $\Lambda$CDM that can be interpreted as footprints of dynamical DE.
%%%%%%%%%%%%%%%%%%%%%%%%%%%%%%%%%%%%%%%%%%%%%

%%%%%%%%%%%%%%%%%%%%%%%%%%%%%%%%%%%%%%%%%%%%%

\section{Methodology}
We construct a generalized Padé-$(2,1)$ expansion around an arbitrary pivot redshift $z_0 \ne 0$. 
This formulation expresses all relevant observables directly in terms of the cosmographic parameters $q(z_0)$, $j(z_0)$, and $s(z_0)$ at the chosen pivot. The explicit derivation is provided in \hyperref[section:A]{Appendix A}.

Generalizing the Padé-$(2,1)$ expansion to an arbitrary pivot redshift $z_0$ offers several key advantages. Firstly, because the accuracy of any truncated expansion inevitably deteriorates with distance from the expansion center, our framework allows the pivot redshift to be chosen strategically so as to minimize truncation errors across the redshift range populated by low-$z$ data. To quantify this effect, we fix the cosmographic parameters to their $\Lambda$CDM values and compare Padé approximations of the key observables with the exact $\Lambda$CDM functions for different choices of $z_0$. When expanded around $z_0 = 0$, the Padé-$(2,1)$ already produces theoretical uncertainties comparable to current DESI BAO error bars. By contrast, for higher pivot redshifts, $0.4 \lesssim z_0 \lesssim 1$, the accuracy of the reconstructed observables improves by up to two orders of magnitude across the full data range. In this regime, the generalized Padé-$(2,1)$ reproduces $\Lambda$CDM predictions with negligible relative error, while avoiding the need for higher-order expansions and the additional free parameters they entail.
 
Secondly, in our approach we expand the Hubble rate $H(z)$ and the luminosity distance $D_{\rm L}(z)$ using the Padé framework, and from these two quantities we obtain consistent expressions for all other observables in terms of the cosmographic parameters ${H(z_0), q(z_0), j(z_0), s(z_0)}$ at the chosen pivot $z_0$. By selecting $z_0 \in [0\, ,\,1]$ and fitting predictions to the data, these parameters are constrained exactly where most observations lie, without the need for extrapolation into regions with no data. This contrasts with the standard approach, which fixes parameters at $z=0$ and extrapolates across the entire data range.

Finally, building on the previous point, our framework allows us to choose multiple pivot redshifts $z_0$, yielding constraints on the cosmographic parameters $H(z_0)$, $q(z_0)$, $j(z_0)$, and $s(z_0)$ at different pivots. For quantities with direct physical interpretation such as $H(z_0)$ and $q(z_0)$, this provides a direct mapping of the expansion rate and the deceleration of the Universe as a function of redshift -- not through extrapolation, but via direct measurements at multiple redshift bins. In practice, this delivers a precise trend of their redshift evolution, offering a powerful, model-independent probe of cosmic acceleration. For similar pioneering discussions see also ~\citet{Liu:2023ctw}.

To capitalize on the advantages of our generalized cosmographic framework, we develop a dedicated numerical package fully interfaced with the \texttt{Cobaya} sampler~\citep{Torrado:2020dgo}, which we use to perform parameter inference via Markov Chain Monte Carlo (MCMC) sampling. In practice, we explore $11$ pivot scenarios, fixing $z_0$ to values uniformly distributed between $z_0=0$ and $z_0=1$ in steps of $\Delta z_0=0.1$. For each choice of pivot, we run full MCMC chains to constrain the cosmographic parameters $\{H(z_0), q(z_0), j(z_0), s(z_0)\}$ using different combinations of the most up-to-date cosmological datasets.

Our baseline dataset includes the second DESI BAO data release~\citep{DESI:2025zgx}, calibrated with the Planck value of the sound horizon $r_d=(147.09\pm0.26)$ Mpc~\citep{Planck:2018vyg}, distance moduli from SNIa (both from the PantheonPlus (PP) catalog~\citep{Brout:2022vxf} and the five-year Dark Energy Survey Supernova Program (DESy5)~\citep{DES:2024jxu} sample), and 15 direct measurements of $H(z)$ from Cosmic Chronometers (CC) with a full covariance matrix~\citep{Jimenez:2001gg, Borghi:2021rft}. A more detailed discussion of the methodology, datasets, and precision tests of the generalized Padé expansion is provided in \hyperref[section:B]{Appendix B}.

%%%%%%%%%%%%%%%%%%%%%%%%%%%%%%%%%%%%%%%%%%%%%
\section{Results}

\begin{figure*}[htp!]
    \centering
    \includegraphics[width=\linewidth]{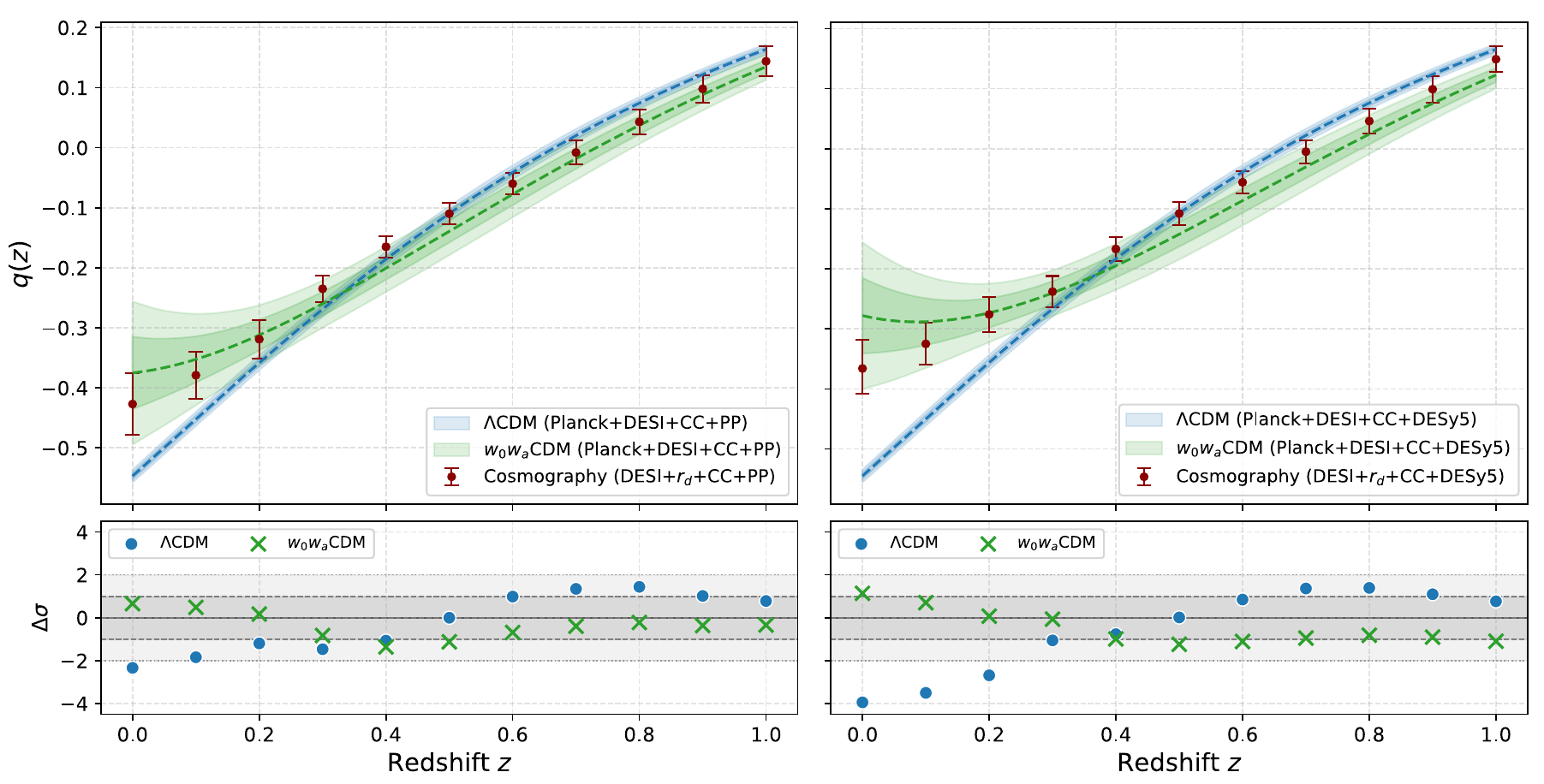}
    \caption{Cosmographic constraints on the deceleration parameter $q(z)$ at different redshift bins with 68\% CL uncertainties (red points). The blue band shows the $\Lambda$CDM predictions, while the green band corresponds to those from a CPL model of dynamical DE, referred to as $w_0w_a$CDM. The left panel shows results obtained by combining Planck (or Planck $r_d$), DESI, and CC with PP SNIa. The right panel shows results from the same data combination but with DESy5 instead of PP. The bottom panels display the distance of the cosmographic points from the model predictions in units of the combined uncertainty $\Delta\sigma$. Blue circles correspond to $\Lambda$CDM and green crosses to $w_0w_a$CDM, providing a direct measure of the level of agreement between the cosmographic constraints and those derived within the respective cosmological models.}
    \label{fig:1}
\end{figure*}

The full numerical results for all dataset combinations are reported in Tab.~\ref{tab:results}, \hyperref[section:C]{Appendix C}. Here we focus on the two most relevant cases: DESI+$r_d$+CC+PP and DESI+$r_d$+CC+DESy5. While these combinations are the most informative in terms of independent probes, we note that, as shown in \hyperref[section:C]{Appendix C}, CC contribute very little statistical weight in these cases, and removing them leaves the cosmographic constraints essentially unchanged.

We begin with the deceleration parameter $q(z)$, which we recall is defined as 
\begin{equation}
q(z) \equiv -\frac{\ddot{a} a}{\dot{a}^2} = \frac{d \ln H}{d \ln (1+z)} - 1 .
\end{equation}
The sign of $q(z)$ indicates whether the Universe is accelerating ($q(z)<0$) or decelerating ($q(z)>0$) at a given redshift, while its value reflects how the expansion rate changes with time.

\begin{figure}[ht!]
    \centering
    \includegraphics[width=\columnwidth]{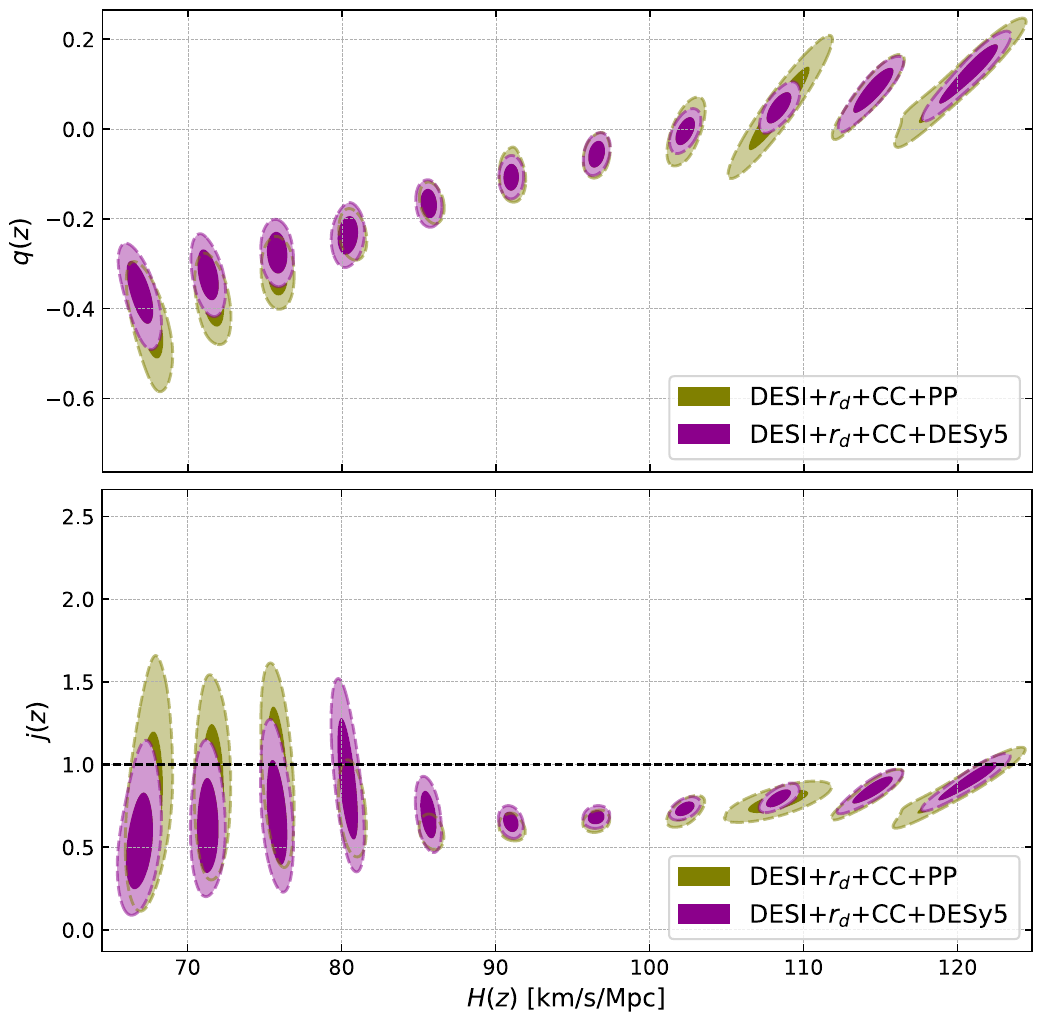}
    \caption{2D correlations between the constraints on the expansion rate, the deceleration parameter, and the jerk parameter obtained at 11 different pivot redshifts, from $z_0=0$ (far left) to $z_0=1$ (far right).}
    \label{fig:2}
\end{figure}

The constraints on $q(z_0)$ obtained with the Padé-$(2,1)$ expansion at different pivot redshifts are shown in Fig.~\ref{fig:1} as red points with their 68\% confidence level (CL) error bars. As seen in the figure, for both dataset combinations the error bars on $q(z_0)$ are largest when the expansion is centered at $z_0=0$ (standard approach) and shrink markedly when the pivot is shifted to intermediate redshifts, $z_0\simeq0.3-0.6$. This behavior has two main origins. First, expansions centered at higher $z_0$ provide a more accurate reconstruction across the redshift range covered by the data, improving the theoretical precision of the series by up to two orders of magnitude compared to $z_0=0$. Second, changing the pivot rotates the correlation between cosmographic parameters. As seen in Fig.~\ref{fig:2}, at $z_0\simeq0$ a strong degeneracy exists between $q(z_0)$ and $H(z_0)$, with more negative values of the former driving larger values of the latter. As the pivot moves to $z_0\simeq0.3-0.6$ this correlation is progressively reduced and eventually lost, yielding much tighter constraints.

This happens because that redshift window is populated by both distance data (DESI, PP/DESy5) and direct $H(z)$ measurements (DESI, CC), making it far easier to disentangle their effects. At higher pivots ($z_0\simeq0.7-1$), however, the correlation between $q(z_0)$ and $H(z_0)$ re-emerges with opposite sign, and uncertainties increase again. Overall, fixing the pivot around $z_0\sim0.3-0.6$ provides the optimal balance: it maximizes the theoretical accuracy of the truncated series while yielding significantly tighter constraints thanks to reduced degeneracies. Additional validation of these statements is provided in \hyperref[section:C]{Appendic C}. We stress that these results are crucial: given the increasing precision of present and future background probes, our generalized framework ensures that theoretical errors remain smaller than observational uncertainties without requiring higher-order expansions or introducing new parameters. This allows cosmographic tests to fully exploit the constraining power of current and forthcoming surveys.

That being said, the most critical information lies in the distribution of the red points in Fig.~\ref{fig:1}. By constraining $q(z)$ at different pivots with high precision, we obtain snapshots of the deceleration parameter at multiple epochs. This provides a direct reconstruction of its redshift evolution, derived purely from cosmography without assuming any underlying model. By contrast, within a cosmological framework such as $\Lambda$CDM or CPL, once the free parameters are constrained the behavior of $q(z)$ is fixed by the model itself.

In Fig.~\ref{fig:1}, the red points from our generalized cosmographic expansion are compared with the blue band, which represents $\Lambda$CDM predictions (68\% and 95\% CL) obtained for the same datasets (Planck+DESI+CC+PP in the left panel and Planck+DESI+CC+DESy5 in the right panel). The bottom panels quantify these differences by showing the distance of the cosmographic points from the model predictions in units of the combined uncertainty $\Delta\sigma$. Across the full redshift range, the cosmographic reconstruction departs from the $\Lambda$CDM prediction in a coherent, redshift-dependent way. At very low redshift ($z\lesssim0.2$), the mismatch is sharp: with DESy5, the red points lie $\sim 4\sigma$ above $\Lambda$CDM, while with PP the offset is smaller but still close to $2\sigma$. At intermediate redshifts ($0.3\lesssim z \lesssim0.5$), the agreement improves, with the cosmographic points falling within $1$–$2\sigma$ of the model. At higher redshifts ($z\gtrsim0.5$), however, the trend reverses, with all cosmographic determinations lying systematically below the $\Lambda$CDM curve. Although each individual deviation remains at the $1.5$–$2\sigma$ level, their persistence across consecutive bins makes the effect statistically meaningful: the cosmographic constraints on $q(z)$ do not simply scatter around the model expectation but reveal a structured, redshift-dependent trend.

This behavior is what one would expect if the late-time expansion of the Universe were driven by a dynamical DE component whose EoS evolves from a present-day quintessence phase ($w(z)>-1$) across the phantom divide ($w(z)<-1$). At low redshift, the cosmographic reconstruction shows $q(z)$ values higher than $\Lambda$CDM, indicating a weaker acceleration. This is consistent with a quintessence-like regime, where the negative pressure of DE is not strong enough to produce the same level of acceleration as a cosmological constant. At intermediate redshift, the quintessence-to-phantom transition occurs, so that around $w(z)\simeq -1$ the deceleration parameter approaches the $\Lambda$CDM prediction. Within benchmark Dynamical DE scenarios, for the datasets analyzed in this work, the phantom crossing is estimated to occur around $z\sim0.3$–$0.4$~\citep{Giare:2024gpk,Ozulker:2025ehg}, precisely where we recover improved consistency with the $\Lambda$CDM predictions in Fig.~\ref{fig:1}. Finally, at higher redshifts, after crossing into the phantom regime, the trend reverses: the negative pressure becomes more repulsive than that of a cosmological constant, the expansion is driven faster than in $\Lambda$CDM, and $q(z)$ falls systematically below the $\Lambda$CDM curve.

To corroborate this hypothesis, in Fig.~\ref{fig:1} we compare the cosmographic reconstruction with the green bands, which show the predictions (68\% and 95\% CL) obtained within the CPL parametrization of the DE EoS for the same datasets. The agreement with the red cosmographic points is remarkable. At low redshift, CPL successfully reproduces the $\gtrsim 2$–$4\sigma$ departures from $\Lambda$CDM, remaining fully consistent with the cosmographic results within $1\sigma$. This lends weight to the interpretation that the observed departures from $\Lambda$CDM may represent footprints of dynamical DE in the late-time expansion history of the Universe.

\begin{figure*}
    \centering
    \includegraphics[width=\linewidth]{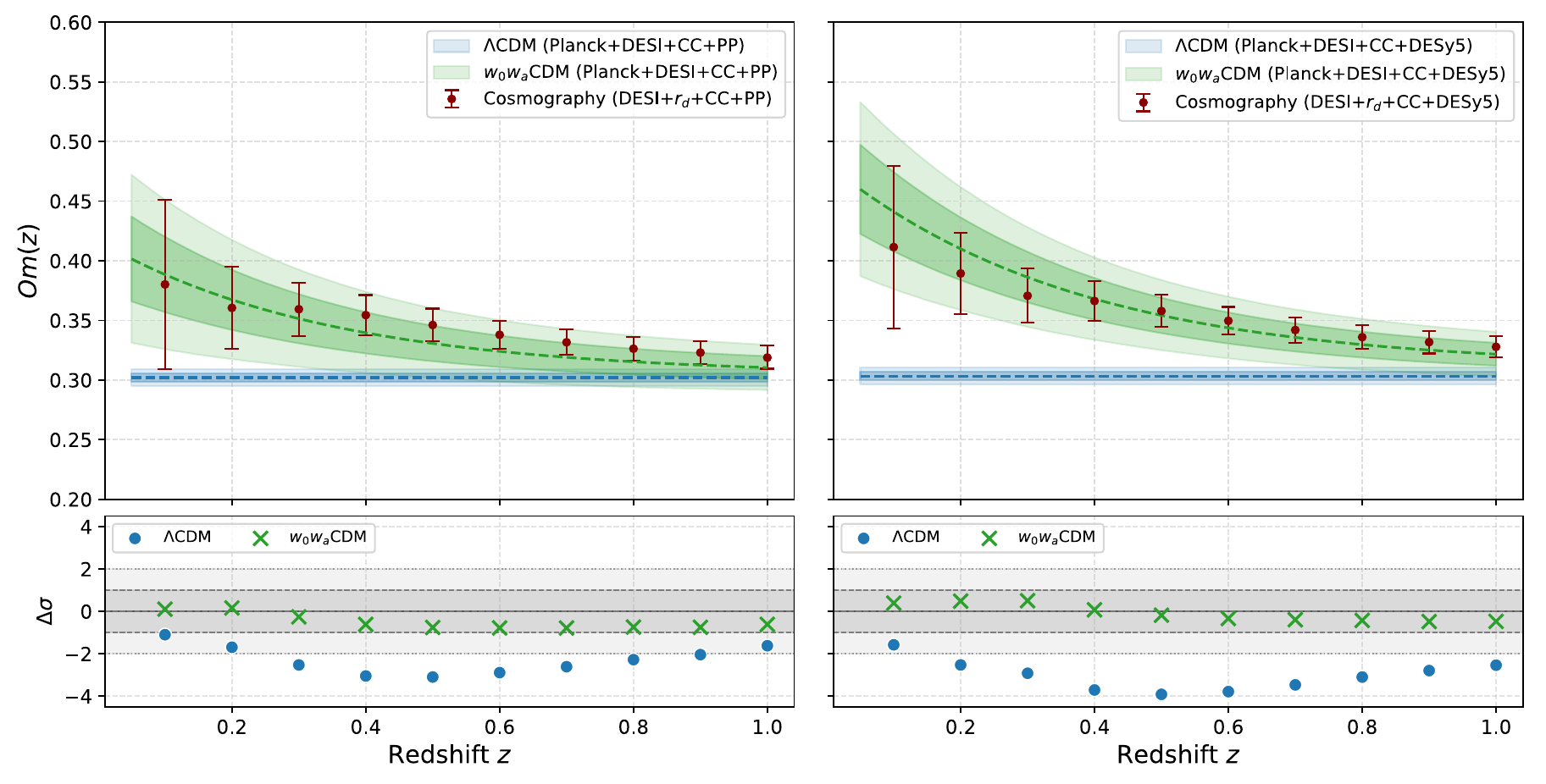}
    \caption{Cosmographic constraints on the $Om(z)$ diagnostic at different redshift bins with 68\% CL uncertainties (red points). The blue band shows the $\Lambda$CDM predictions ($Om(z)=\Omega_m$), while the green band corresponds to those from a CPL model of dynamical DE, referred to as $w_0w_a$CDM. The left panel shows results obtained by combining Planck (or Planck $r_d$), DESI, and CC with PP SNIa. The right panel shows results from the same data combination but with DESy5 instead of PP. The bottom panels display the distance of the cosmographic points from the model predictions in units of the combined uncertainty $\Delta\sigma$. Blue circles correspond to $\Lambda$CDM and green crosses to $w_0w_a$CDM, providing a direct measure of the level of agreement between the cosmographic constraints and those derived within the respective cosmological models.}
    \label{fig:3}
\end{figure*}

To further validate our interpretation, we generalize the $Om(z)$ diagnostic. Following ~\citet{Sahni:2008xx}, we define
\begin{equation}
Om(z) \equiv \frac{\tilde{h}^2(z)-1}{(1+z)^3-1},
\end{equation}
where $\tilde{h}(z)=H(z)/H_0$ is the Hubble parameter normalized to its present value. This diagnostic has a distinctive property: if DE is a cosmological constant, $Om(z)$ remains exactly constant at all redshifts, with $Om(z)=\Omega_m$. Any dynamical behavior in the DE sector instead produces a clear redshift dependence. For instance, as pointed out in Fig.~3 of ~\citet{Sahni:2008xx}, a descending trend in $Om(z)$ should be observed for a quintessence-like DE EoS. Our method, which constrains $H(z)$ cosmographically at multiple pivots, delivers precise determinations of $H(z)$ across the redshift range. This enables us to constrain $Om(z)$ at several redshifts and directly test whether it is consistent with a constant value or reveals a redshift-dependent trend.

The results are shown in Fig.~\ref{fig:3}, where the red points represent our cosmographic determinations of $Om(z)$ (the left panel corresponds to combinations including PP, the right panel to those including DESy5). The red points are far from constant and deviate significantly from the $\Lambda$CDM prediction shown in blue. In both datasets, the discrepancy with $\Lambda$CDM emerges already at $z\gtrsim0.1$ and persists across the entire redshift range. At low redshift, the uncertainties in $Om(z)$ are larger for the same reasons discussed for $q(z)$, but the trend is nonetheless visible. At intermediate redshifts, the benefit of our generalized cosmographic method becomes evident: the constraints are significantly tighter, and the red points remain systematically above the $\Lambda$CDM curve. In the bottom panels we quantify these differences by showing the distance of the cosmographic points from the model predictions in units of the combined uncertainty $\Delta\sigma$. Depending on the dataset, the cosmographic estimates of $Om(z)$ at $z\gtrsim0.3$ exceed the $\Lambda$CDM prediction by $3$–$3.5\sigma$ for combinations including PP and by as much as $\sim4\sigma$ for DESy5. At higher redshift the trend continues, with all determinations consistently displaced from the $\Lambda$CDM expectation.

As in the case of $q(z)$, our determinations of $Om(z)$ do not scatter randomly around the $\Lambda$CDM prediction but reveal a coherent redshift-dependent trend. This becomes especially clear when comparing the red points with the green bands, which represent the CPL predictions for the same datasets. In this case, the agreement with dynamical DE models is remarkable: the red cosmographic points follow the CPL best-fit predictions (dotted green curve) almost perfectly across the full redshift range, both for Planck+DESI+CC+PP and Planck+DESI+CC+DESy5. This demonstrates that the cosmographic expansion, when translated into $Om(z)$ constraints, yields results in excellent agreement with dynamical DE (already within a simple CPL description) rather than with a cosmological constant.

Beyond $q(z)$ and $Om(z)$, further cosmographic parameters such as $H(z)$ and $j(z)$ also show systematic departures from $\Lambda$CDM, consistently pointing toward dynamical DE. In particular, our constraints on $j(z)$ reveal significant departures from the $\Lambda$CDM identity $j(z)\equiv 1$, see Fig.~\ref{fig:2}. Similarly, by constraining the expansion rate $H(z)$ at different pivot redshifts, our methodology indicates a preference for a faster expansion at $z\gtrsim0.3$, about $2$ km/s/Mpc above the $\Lambda$CDM prediction, while coming into closer agreement at lower redshift ($z\lesssim0.3$). At $z=0$, the Padé series yields $H_0 = 67.49 \pm 0.56$ ($66.90 \pm 0.53$) km/s/Mpc for DESI+$r_d$+PP+CC (DESI+$r_d$+DESy5+CC). These values are lower than the $H_0$ inferred from the same datasets under the assumption of $\Lambda$CDM by roughly 1 km/s/Mpc, while remaining fully consistent with the values obtained assuming a CPL dynamical DE model. A more detailed discussion of these results is presented in \hyperref[section:C]{Appendix C}. We also note that the absolute normalization of $H(z)$ depends on the calibration adopted for BAO, which in our baseline analysis is anchored to the Planck determination of the sound horizon within $\Lambda$CDM. In \hyperref[section:D]{Appendix D} we present an extended analysis exploring alternative calibrations of the datasets (e.g., SH0ES–based calibration of SNIa or reduced-$r_d$ priors motivated by models of interest for the Hubble tension), showing that such choices simply rescale $H(z)$ while leaving the cosmographic constraints on $q(z)$ and $Om(z)$ -- and thus our main conclusions -- completely unchanged.

Overall, the cosmographic reconstructions at multiple pivots are consistently better matched by simple dynamical DE models such as CPL, providing an additional and independent indication of DE dynamics and offering a clear, model-independent guideline for building more successful theoretical scenarios to interpret current data.

%%%%%%%%%%%%%%%%%%%%%%%%%%%%%%%%%%%%%%%%%%%%%
\section{Conclusions}

In this work we have developed a generalized cosmographic framework based on Padé-$(2,1)$ expansions around arbitrary pivot redshifts $z_0$, showing that it overcomes several intrinsic limitations of the standard approach, which relies on truncated Taylor or Padé series around $z_0=0$. Our method provides three key methodological advantages:

\begin{itemize}
\item \textbf{\textit{Accuracy:}} Expanding the Padé-$(2,1)$ series around higher pivots reduces truncation errors across the full redshift range probed by current data ($0\lesssim z \lesssim 2.5$), improving the accuracy of reconstructed observables by up to two orders of magnitude.
\item \textbf{\textit{Precision:}} Expressing all observables directly in terms of cosmographic parameters defined at the chosen pivot ensures that parameters are constrained exactly where data are available. This removes the need for extrapolation, reduces degeneracies, and allows for a much cleaner disentangling of their effects, resulting in significantly more precise constraints.
\item \textbf{\textit{Flexibility:}} By repeating the analysis at multiple pivots we obtain high-precision constraints on cosmographic parameters at different epochs, directly mapping the redshift evolution of key quantities such as $H(z)$ and $q(z)$. In this way, cosmography becomes a precise, model-independent tracer of the late-time kinematics of the Universe.
\end{itemize}

To exploit these advantages, we derive cosmographic constraints at 11 pivot redshifts uniformly distributed between $z_0=0$ and $z_0=1$. We analyze the latest DESI BAO measurements calibrated with the $\Lambda$CDM Planck value of the sound horizon, SNIa from PP and DESy5, and Cosmic Chronometers. The main results can be summarized as follows:

\begin{itemize}
\item \textbf{\textit{Pivot redshift and correlations:}} As seen in Fig.~\ref{fig:2}, at $z_0\simeq0$, $H(z_0)$ and $q(z_0)$ are strongly anticorrelated, and the uncertainties on their inferred values are large. Moving $z_0$ to higher redshift, the degeneracy line rotates so that around $z_0\simeq0.3-0.6$ the two parameters become nearly decorrelated and their effects can be cleanly disentangled, yielding significantly tighter constraints. At still higher pivots ($z_0\simeq0.7-1$), the correlation continues to rotate and reappears with opposite sign, again enlarging the inferred uncertainties. With current data, the optimal pivot range is therefore $z_0\simeq0.3-0.6$.
\item \textbf{\textit{Footprints of Dynamical DE in $\boldsymbol{q(z)}$:}} The cosmographic constraints on the deceleration parameter shown in Fig.~\ref{fig:1} deviate from the $\Lambda$CDM predictions in a coherent, redshift-dependent way. At $z\lesssim0.2$, $q(z)$ lies significantly above the $\Lambda$CDM curve (up to $\sim4\sigma$ for DESy5 and $\sim2\sigma$ for PP), while at $z\gtrsim0.6$ it falls systematically below it. Overall, the reconstructed trend shows much closer agreement with the behavior predicted within a CPL parametrization of DE.
\item \textbf{\textit{Footprints of Dynamical DE in $\boldsymbol{Om(z)}$:}} We generalize the $Om(z)$ diagnostic to constrain this quantity at different pivot redshifts. The cosmographic constraints shown in Fig.~\ref{fig:3} are inconsistent with the constant behavior expected in $\Lambda$CDM, as they lie systematically above the $\Lambda$CDM curve (with deviations of $2.5$–$3.5\sigma$ for combinations including PP and up to $\sim4\sigma$ for DESy5), showing a redshift trend that aligns almost perfectly with the CPL best-fit predictions across the entire redshift range.
\end{itemize}

Taken together, these results provide a coherent and assumption-free indication of deviations from $\Lambda$CDM and suggest that we are observing the footprints of dynamical DE in the late-time kinematics of the Universe. Additional cosmographic constraints reinforce this interpretation: the jerk parameter $j(z)$ departs from the $\Lambda$CDM identity $j\equiv1$, while the expansion rate $H(z)$ is faster than $\Lambda$CDM at $z\gtrsim0.3$ by about $2$ km/s/Mpc, in close alignment with CPL predictions. More broadly, since our constraints on the cosmographic quantities are by definition independent of the underlying dynamics, they also represent universal consistency conditions that any model of cosmic acceleration should satisfy.

Looking ahead, our generalized cosmographic framework enables high-precision, model-independent tests of cosmic acceleration. With forthcoming data from DESI and Euclid, it will be crucial to keep theoretical truncation errors below observational uncertainties. Our approach ensures higher accuracy without higher-order expansions, reduces parameter degeneracies, and provides a robust and flexible tool to probe cosmic acceleration.
%%%%%%%%%%%%%%%%%%%%%%%%%%%%%%%%%%%%%%%%%%%%%

\begin{acknowledgments}
EF is supported by the research grant number 2022E2J4RK “PANTHEON: Perspectives in Astroparticle and Neutrino THEory with Old and New messengers” under the program PRIN 2022 funded by the Italian Ministero dell’Università e della Ricerca (MUR). 
WG is supported by the Lancaster–Sheffield Consortium for Fundamental Physics under STFC grant: ST/X000621/1.
EDV is supported by a Royal Society Dorothy Hodgkin Research Fellowship. This article is based upon work from COST Action CA21136 Addressing observational tensions in cosmology with systematics and fundamental physics (CosmoVerse) supported by COST (European Cooperation in Science and Technology). We acknowledge IT Services at The University of Sheffield for the provision of services for High Performance Computing.
\end{acknowledgments}

\begin{contribution}
EF derived the generalized cosmographic framework used in this work, implemented it numerically, performed the MCMC analyses, and contributed to editing the manuscript.
WG conceived the initial research idea, supervised the theoretical derivation, supervised and contributed to the numerical implementation of the cosmographic expansion within the cosmological pipeline, and wrote the first draft of the manuscript.
EDV validated both the theoretical and numerical analyses and contributed to editing the manuscript.
All authors contributed to refining and finalizing the manuscript.
\end{contribution}

\clearpage

\appendix

\section{Generalized Cosmographic expansion} 
\label{section:A}
Cosmography describes the late-time expansion history of the Universe by expressing cosmological observables as series expansions.\footnote{With no claim of completeness regarding the plethora of works employing cosmographic analyses over the last couple of decades to probe cosmological properties, we refer the reader to~\citet{Saini:1999ba,Bernstein:2003es,John:2005bz,Cattoen:2007id,Capozziello:2008tc,Xu:2010hq,Luongo:2011zz,Carvalho:2011qw,Demianski:2012ra,Bamba:2012cp,Lazkoz:2013ija,Springob:2014qja,Gao:2014iva,Jee:2015yra,Zhou:2016nik,Demianski:2016dsa,Zhang:2016urt,Martins:2016bbi,Shajib:2017omw,Yang:2019vgk,Li:2019qic,Benetti:2019gmo,Escamilla-Rivera:2019aol,Yin:2019rgm,Rezaei:2020lfy,Chen:2020knz,Banerjee:2020bjq,Escamilla-Rivera:2021vyw,Treu:2022aqp,Hu:2022udt,Pourojaghi:2022zrh,Rocha:2022gog,Pourojaghi:2024bxa,Jesus:2024nrl,Rodrigues:2025tfg,Yang:2025jei,Ling:2025lmw}.} The standard approach employs a Taylor expansion in redshift $z$, centered on the present epoch $z_0=0$. In this framework, observables such as the luminosity distance $D_{\rm L}(z)$ and the expansion rate $H(z)$ can be written in terms of cosmographic parameters defined at $z=0$~\citep{Visser:2003vq, Visser:2004bf}. Constraining these parameters from data then determines the behavior of the series at any redshift, providing a model-independent description of the Universe’s kinematics.\footnote{The nature of cosmic acceleration, and in particular the question of whether it is driven by a cosmological constant or emerges from dynamical models, has long attracted significant attention in the high-energy community, even before the release of DESI collaboration data. For pioneering works in this direction, we refer to ~\citet{Cooray:1999da,Efstathiou:1999tm,Chevallier:2000qy,Linder:2002et,Wetterich:2004pv,Feng:2004ff,Hannestad:2004cb,Xia:2004rw,Gong:2005de,Jassal:2005qc,Nesseris:2005ur,Liu:2008vy,Barboza:2008rh,Barboza:2009ks,Ma:2011nc,Sendra:2011pt,Feng:2011zzo,Barboza:2011gd,DeFelice:2012vd,Feng:2012gf,Wei:2013jya,Magana:2014voa,Akarsu:2015yea,Pan:2016jli,DiValentino:2016hlg,Nunes:2016plz,Nunes:2016drj,Magana:2017usz,Yang:2017alx,Pan:2017zoh,Panotopoulos:2018sso,Yang:2018qmz,Jaime:2018ftn,Das:2017gjj,Yang:2018prh,Li:2019yem,Yang:2019jwn,Pan:2019hac,Tamayo:2019gqj,Pan:2019brc,DiValentino:2020naf,Rezaei:2020mrj,Perkovic:2020mph,Banihashemi:2020wtb,Jaber-Bravo:2019nrk,Benaoum:2020qsi,Yang:2021eud,Jaber:2021hho,Alestas:2021luu,Yang:2022klj,Escudero:2022rbq,Castillo-Santos:2022yoi,Yang:2022kho,Dahmani:2023bsb,Escamilla:2023oce,Rezaei:2023xkj,Adil:2023exv,LozanoTorres:2024tnt,Singh:2023ryd,Rezaei:2024vtg,Reyhani:2024cnr}.} 

However, while the Taylor expansion performs well at low redshifts, it rapidly loses accuracy at higher redshifts, where it may even diverge, introducing biases and large truncation errors. One may attempt to improve the approximation by including higher-order terms, but any expansion in $z$ with $a=1/(1+z)$ has a singularity at $z=-1$~\citep{Cattoen:2007sk, OColgain:2021pyh}, which fixes the radius of convergence to $|z|\le1$ according to the Cauchy-Hadamard theorem. As a result, adding more terms does not necessarily extend the validity of the expansion and may even worsen it beyond $z\simeq1$. This limits the applicability of standard cosmographic expansions in the regime $z \gtrsim 1$, precisely where recent datasets such as DESI BAO and Type Ia supernovae (SNIa) provide the most constraining power. A common workaround is to reformulate the expansion in terms of the auxiliary variable $y=z/(1+z)$, which removes the singularity and ensures convergence over the full redshift range. However, the $y$ variable compresses the redshift interval and acts as a smaller expansion parameter, leading to larger propagated uncertainties and requiring more coefficients to achieve the same accuracy~\citep{Busti:2015xqa}.

To address the shortcomings of Taylor series, Padé approximants have been proposed, replacing the polynomial expansion with a ratio of polynomials~\citep{Gruber:2013wua, Wei:2013jya, Aviles:2014rma}. Among them, the Padé-$(2,1)$ expansion has been shown to offer the best compromise between accuracy, stability, and the number of cosmographic parameters involved~\citep{Capozziello:2018jya, Capozziello:2020ctn, Hu:2022udt, Yu:2025ytv}. 

Despite these improvements, significant limitations remain. In its standard formulation, the Padé expansion is still built from truncated series around $z_0=0$, so the fitted parameters characterize only the present epoch, while the behavior at $z>0$ is inferred indirectly from the series itself. As redshift increases, the approximation still deteriorates and truncation errors can reach or even exceed observational uncertainties precisely in the range where most data are available ($0.1 \lesssim z \lesssim 2$). Extending the series to higher orders can mitigate this, but at the cost of introducing extra parameters, which increases degeneracies and undermines predictivity.

\clearpage
To overcome these limitations, in this work we refine the cosmographic approach by formulating the Padé-$(2,1)$ expansion around a generic pivot redshift $z_0$. This enables us to fit cosmological data at higher redshifts with improved precision and to constrain cosmographic parameters directly at different redshift values. In turn, this allows us to reconstruct the full cosmographic functions, which can then be employed to test the nature of cosmic acceleration. \footnote{Following the DESI results, this interest has been widely revitalized, with a multitude of analyses and reinterpretations of the observations. For a non-complete yet broad survey of the existing literature in this direction, see for instance ~\citet{Cortes:2024lgw,Shlivko:2024llw,Luongo:2024fww,Yin:2024hba,Gialamas:2024lyw,Dinda:2024kjf,Najafi:2024qzm,Wang:2024dka,Ye:2024ywg,Tada:2024znt,Carloni:2024zpl,Chan-GyungPark:2024mlx,DESI:2024kob,Bhattacharya:2024hep,Ramadan:2024kmn,Notari:2024rti,Orchard:2024bve,Hernandez-Almada:2024ost,Pourojaghi:2024tmw,Giare:2024gpk,Reboucas:2024smm,Giare:2024ocw,Chan-GyungPark:2024brx,Menci:2024hop,Li:2024qus,Li:2024hrv,Li:2024qso,Notari:2024zmi,Gao:2024ily,Fikri:2024klc,Jiang:2024xnu,Zheng:2024qzi,Gomez-Valent:2024ejh,RoyChoudhury:2024wri,Lewis:2024cqj,Wolf:2025jlc,Shajib:2025tpd,Giare:2025pzu,Chaussidon:2025npr,Kessler:2025kju,Pang:2025lvh,RoyChoudhury:2025dhe,Scherer:2025esj,Teixeira:2025czm,Specogna:2025guo,Cheng:2025lod,Cheng:2025hug,Ozulker:2025ehg,Lee:2025pzo,Silva:2025twg,Ishak:2025cay,Li:2025owk,Lu:2025gki,Ling:2025lmw}.}

In this section, we derive, step by step, the generalized cosmographic expansion (Sec.~\ref{sec.Pade.derivation}) and show that our formulation reproduces the known results when $z_0=0$ (Sec.~\ref{sec.Pade.test}).

\subsection{Derivation of Padé-$(2,1)$ expansion around a pivot redshift $z_0$}
\label{sec.Pade.derivation}

In general mathematical terms, the Padé-$(2,1)$ expansion of a function is defined as~\citep{Gruber:2013wua, Aviles:2014rma} 
\begin{equation}
f^{(2,1)}(z)=\frac{a_0+a_1x+a_2x^{2}}{1+b_1x},
\label{eq:pade_definition_2}
\end{equation}
where $a_i$ and $b_i$ are the Padé coefficients. The Padé approximants are degenerate under a rescaling factor; for this reason $b_0$ is usually fixed to $b_0=1$.

To express these coefficients explicitly in terms of cosmographic parameters, we follow the standard procedure of comparing the Taylor expansion with the Padé form. The first step is to expand the denominator of the Padé expression in Eq.~\eqref{eq:pade_definition_2} as a Taylor series, obtaining
\begin{equation}
f^{(2,1)}(z)=
a_0+\bigl(a_1-a_0b_1\bigr)x+\bigl(a_2-a_1b_1+a_0b_1^{2}\bigr)x^{2}
+\bigl(-a_2b_1+a_1b_1^{2}-a_0b_1^{3}\bigr)x^{3}+{\cal O}(x^{4}).
\label{eq:Pade_expansion}
\end{equation}
In this way the Padé-$(2,1)$ expression can be compared with the third-order Taylor expansion of a function, written in general as
\begin{equation}
f^{\mathrm{Tay}}(z)=c_0+c_1x+c_2x^{2}+c_3x^{3}+{\cal O}(x^{4}),
\label{eq:Tayalor_ci}
\end{equation}
Equating term by term Eq.~\eqref{eq:Tayalor_ci} and Eq.~\eqref{eq:Pade_expansion}, we can fix the Padé parameters:
\begin{equation}
\begin{aligned}
c_0 &= a_0,\\
c_1 &= a_1-a_0b_1,\\
c_2 &= a_2-a_1b_1+a_0b_1^{2},\\
c_3 &= -a_2b_1+a_1b_1^{2}-a_0b_1^{3}.
\end{aligned}
\end{equation}
Solving this triangular system, we finally obtain
\begin{equation}
a_0=c_0,\qquad
a_1=c_1+c_0b_1,\qquad
a_2=c_2+c_1b_1,\qquad
b_1=-\frac{c_3}{c_2}.
\label{eq:Pade_coeff}
\end{equation}
In the following section, we apply this methodology to derive a generalized Padé-$(2,1)$ expansion centered around a generic redshift. We proceed as follows. First, in Sec.~\ref{subsec.Taylor.H0} we present the derivation of a third-order Taylor expansion for the Hubble parameter $H(z)$ around an arbitrary pivot redshift $z=z_0$. In Sec.~\ref{subsec.Taylor.DL}, we present the derivation of a third-order Taylor expansion for the luminosity distance $D_L(z)$ around an arbitrary pivot redshift $z=z_0$. Building on this, we then construct the Padé-$(2,1)$ approximation for both the luminosity distance (Sec.~\ref{subsec.Pade.DL}) and the expansion rate (Sec.~\ref{subsec.Pade.H}) by adopting the definition of the Padé-$(2,1)$ series and matching terms order by order to determine the Padé coefficients. 

\subsubsection{Taylor Expansion of the Hubble parameter around a pivot redshift $z_0$}
\label{subsec.Taylor.H0}

Here, we present the derivation of a third-order Taylor expansion for the Hubble parameter $H(z)$ and the luminosity distance $D_L(z)$ around an arbitrary pivot redshift $z=z_0$. 

We begin with the Taylor expansion of the Hubble parameter, $H(z)^{\mathrm{Tay}}$, around $z=z_0$, truncated at third order, which can be written as:
\begin{equation}
    H(z)^{\mathrm{Tay}} = H(z_0) + H'(z_0)(z-z_0) + \frac{1}{2}\,H''(z_0)(z-z_0)^2 + \frac{1}{6}\,H'''(z_0)(z-z_0)^3 + \mathcal{O}\bigl((z-z_0)^4\bigr).
    \label{eq:app1}
\end{equation}
It is convenient to factor out $H(z_0)$ and write
\begin{equation}
    H^{\mathrm{Tay}}(z)= H_{z_0}\left[ 1 + Ax + Bx^2 + Cx^3 + \mathcal{O}\bigl(x^4\bigr) \right],
    \label{eq:H_def}
\end{equation}
where $x\equiv z-z_0$, we adopt the notation $Y_{z_0}\equiv Y(z_0)$ for any generic function $Y$ evaluated at $z_0$ (so $H_{z_0}\equiv H(z_0)$), and define the dimensionless coefficients
\begin{equation}
    A = \frac{H'_{z_0}}{H_{z_0}}, \quad B = \frac{H''_{z_0}}{2\,H_{z_0}}, \quad C = \frac{H'''_{z_0}}{6\,H_{z_0}}\,.
    \label{eq:app3}
\end{equation}

Our goal is to express these coefficients in terms of cosmographic parameters evaluated at $z_0$. For a third-order expansion, the relevant quantities are the deceleration parameter $q(z)$, the jerk parameter $j(z)$, and the snap parameter $s(z)$, respectively defined as~\citep{Visser:2003vq}:
\begin{equation}
q = - \frac{d^2a/dt^2}{a \, H^2},\qquad
j = \frac{d^3a/dt^3}{a \,H^3},\qquad
s = \frac{d^4a/dt^4}{a \, H^4}\,.
\label{eq:cosmo_par_def}
\end{equation}

To express $A$, $B$, and $C$ in terms of $q_{z_0}$, $j_{z_0}$, and $s_{z_0}$, we first rewrite the derivatives of the scale factor $a(t)$ in terms of derivatives of the Hubble parameter $H(t)$. Starting from the definition of the Hubble parameter $H\equiv\frac{\dot{a}}{a}$ and differentiating it with respect to time, we find
\begin{equation}
    \frac{\ddot{a}}{a}=\dot{H}+H^2.
    \label{eq:app5_rel}
\end{equation}
Using the relation between time and redshift derivatives, $\frac{d}{dt}=-(1+z)H\frac{d}{dz}$, we get
$\dot{H}(t) = -(1+z)\,H(z)\,H'(z)$. Substituting these relations into Eq.~\eqref{eq:cosmo_par_def}, one obtains
\begin{equation}
    q(z) = -1 + \frac{(1+z)H'(z)}{H(z)}.
    \label{eq:qz_def}
\end{equation}
Evaluating this at $z=z_0$, we finally find
\begin{equation}
   A = \frac{H'_{z_0}}{H_{z_0}} = \frac{1+q_{z_0}}{1+z_0}.
    \label{eq:A_def}
\end{equation}

To obtain the jerk parameter we have to reiterate the above process starting from the relation
\begin{equation}
    \frac{d}{dt}\left(\frac{\ddot{a}}{a}\right)=\frac{\dddot{a}}{a}-\frac{\ddot{a}H}{a} \,
    \label{eq:app8}
\end{equation}
and using the Eq.~\eqref{eq:app5_rel} and the definition of the deceleration parameter in Eq.~\eqref{eq:cosmo_par_def} we find
\begin{equation}
    j(t)=\frac{1}{H^3}(\ddot{H}+2H\dot{H})-q(t)\, .
    \label{eq:app9}
\end{equation}
Now, switching to the redshift variable (trough the derivative rule) and rearranging the expression, we obtain
\begin{equation}
    j(z)=q(z)^2+(1+z)^2\frac{H''(z)}{H(z)} \,,
    \label{eq:jz_def}
\end{equation}
from which it is straightforward to find
\begin{equation}
    B=\frac{H''_{z_0}}{2H_{z_0}}=\frac{j_{z_0}-q_{z_0}^2}{2(1+z_0)^2} \,.
    \label{eq:B_def}
\end{equation}

\noindent Lastly, from the definition of the snap parameter in Eq.~\eqref{eq:cosmo_par_def} and from
\begin{equation}
    \frac{d}{dt}\left(\frac{\dddot{a}}{a}\right)=\frac{{a^{(4)}}}{a}-\frac{\dddot{a}H}{a} \,,
    \label{eq:app12}
\end{equation}
and using also the definition of the jerk parameter, we find
\begin{equation}
    s(t)=\frac{1}{H^4}\frac{d}{dt}\left( j(t)H^3\right)+j(t) \,.
    \label{eq:app13}
\end{equation}
Using the definition in Eq.~\eqref{eq:jz_def} and switching to the redshift variable, we obtain
\begin{equation}
    s(z)=-(1+z)\left[2q(z)q'(z)+3q(z)^2\frac{H'}{H}+\frac{H''}{H}+(1+z)\left(\frac{H'''}{H}+2\frac{H'H''}{H^2}  \right)\right]+j(z) \,.
    \label{eq:app14}
\end{equation}
Applying the first derivative to Eq.~\eqref{eq:qz_def} and after some algebraic manipulations, we finally get
\begin{equation}
    s(z)=3q(z)^3+3q(z)^2-4q(z)j(z)-3j(z)-(1+z)^3\frac{H'''}{H} \,,
    \label{eq:sz_def}
\end{equation}
from which it is immediate to obtain 
\begin{equation}
    C=\frac{H'''_{z_0}}{6H_{z_0}}=\frac{3q_{z_0}^3+3q_{z_0}^2-(4q_{z_0}+3)j_{z_0}-s_{z_0}}{6(1+z_0)^3} \,.
    \label{eq:C_def}
\end{equation}

\subsubsection{Taylor Expansion of the Luminosity distance around a pivot redshift $z_0$}
\label{subsec.Taylor.DL}

The luminosity distance in a flat, homogeneous, and isotropic Universe described by the Friedmann-Lemaître-Robertson-Walker (FLRW) metric is defined as
\begin{equation}
    D_{\rm L}(z)=(1+z)\int_0^z\frac{dz'}{H(z')}\,.
    \label{eq:dL_def}
\end{equation}
Because of the inverse Hubble parameter in its formulation, it is inversely proportional to the coefficients $A$, $B$, and $C$. In order to obtain an analytically solvable integral, it is possible to expand $1/H$ as a Taylor series around $z_0$, keeping terms up to second order in $x$ (since we are interested in the luminosity distance up to third order). This yields
\begin{equation}
    \frac{1}{H(z)}=\frac{1}{H_{z_0}}\left(1-Ax+(A^2-B)x^2\right) \,.
    \label{eq:inv_H}
\end{equation}
Integrating this expression leads to the final expression for the Taylor-expanded luminosity distance:
\begin{equation}
D_{\rm L}^{Tay}(z)
= \frac{(1+z)}{H_{z0}}
\left\{ \,z - \frac{A}{2}\,\left[(z-z_0)^2-z_0^2\right] + \frac{A^2-B}{3}\,\left[(z-z_0)^3+z_0^3\right]\right\}\,.
\label{eq:Taylor_final}
\end{equation}

\subsubsection{Padé-$(2,1)$ expansion of the luminosity distance around a pivot redshift $z_0$}
\label{subsec.Pade.DL}

Following the methodology discussed in Sec.~\ref{sec.Pade.derivation}, after deriving the expression for the Taylor series, we can obtain the Padé-$(2,1)$ expansion of the luminosity distance by rewriting the Taylor expansion of $D_L(z)$ in Eq.~\eqref{eq:Taylor_final} as 
\begin{equation}
D_{\rm L}^{\mathrm{Tay}}(z)=c_0+c_1x+c_2x^{2}+c_3x^{3}+{\cal O}(x^{4}),
\label{eq:Tayalor_ci_dl}
\end{equation}
where
\begin{equation}
\begin{aligned}
c_0 &= \frac{(1+z_0)}{H_{z_0}}\left[z_0+\frac{A}{2}z_0^{2}+\frac{(A^{2}-B)}{3}z_0^{3}\right],\\
c_1 &= \frac{1}{H_{z_0}}\left[1+2z_0+\frac{A}{2}z_0^{2}+\frac{(A^{2}-B)}{3}z_0^{3}\right],\\
c_2 &= \frac{1}{2H_{z_0}}\left[2-A(1+z_0)\right],\\
c_3 &= \frac{1}{6H_{z_0}}\left[2(1+z_0)(A^{2}-B)-3A\right].
\end{aligned}
\label{eq:ci_coefficients}
\end{equation}
This directly fixes the Padé coefficients $a_i$ and $b_i$ in terms of the parameters $c_i$ above (and thus in terms of $H_{z_0}$, $q_{z_0}$, $j_{z_0}$, and $s_{z_0}$) by means of Eq.~\eqref{eq:Pade_coeff}.

\subsubsection{Padé-$(2,1)$ expansion of the expansion rate around a pivot redshift $z_0$}
\label{subsec.Pade.H}

To derive the expression of the Padé-$(2,1)$ expansion of the Hubble parameter, we analogously begin from the Taylor expansion of the Hubble parameter in Eq.~\eqref{eq:H_def}, rewritten as 
\begin{equation}
    H^{\mathrm Tay}(z)=c_0+c_1x+c_2x^2+c_3x^3+{\cal O}(x^{4}) \,,
    \label{eq:HTay_Ci}
\end{equation}
where 
\begin{equation}
\begin{aligned}
c_0 &= H_{z_0},\\
c_1 &= H_{z_0}A,\\
c_2 &= H_{z_0}B ,\\
c_3 &= H_{z_0}C .
\end{aligned}
\label{eq:Ci_coefficients}
\end{equation}
Again, this directly fixes the Padé coefficients $a_i$ and $b_i$ in terms of the parameters $c_i$ above (and thus in terms of $H_{z_0}$, $q_{z_0}$, $j_{z_0}$, and $s_{z_0}$) by means of Eq.~\eqref{eq:Pade_coeff}.

\subsection{Consistency Checks}
\label{sec.Pade.test}

As a consistency test, we check that for $z_0=0$ we recover the expressions for the Hubble parameter and luminosity distance corresponding to the standard Taylor and Padé-$(2,1)$ expansions commonly found in the literature (see, e.g., ~\citet{Visser:2004bf}).

We start with the Taylor expansions, $H^{\mathrm{Tay}}$ and $D_{\rm L}^{\mathrm{Tay}}$. Evaluating Eqs.~\eqref{eq:A_def}, \eqref{eq:B_def}, and \eqref{eq:C_def} at $z_0=0$, we obtain
\begin{equation}
H^{\mathrm{Tay}}(z; z_0=0)=H_0\left[1+(1+q_0)z+\frac{j_0-q_0^2}{2}z^2+\frac{3q_0^3+3q_0^2-(4q_0+3)j_0-s_0}{6}z^3\right],
\label{eq:Hubble_z0_Tay}
\end{equation}
\begin{equation}
D_{\rm L}^{\mathrm{Tay}}(z; z_0=0)=\frac{1}{H_0} \left(z+\frac{1-q_0}{2}z^2-\frac{1-q_0-3q_0^2+j_0}{6} z^3\right),
\label{eq:dl_z0_tay}
\end{equation}
which correctly reproduce the results usually found in the literature.

Similarly, it is straightforward to verify that the generalized Padé-$(2,1)$ expansion of the luminosity distance $D_{\rm L}^{\mathrm{P(2,1)}}$ and the expansion rate $H^{\mathrm{P(2,1)}}$ reduces to the known expressions when fixing $z_0=0$. Indeed, computing Eqs.~\eqref{eq:A_def}, \eqref{eq:B_def}, and \eqref{eq:C_def} at $z_0=0$, we find
\begin{equation}
\begin{aligned}
c_0 &= 0, \\
c_1 &= \frac{1}{H_0}, \\ 
c_2 &= \frac{1}{2H_0}(1-q_0),\\ 
c_3 &= -\frac{1}{6H_0}(1-q_0-3q_0^2+j_0),
\end{aligned}
\label{eq:ci_coefficients_z_0}
\end{equation}
which are exactly the Taylor coefficients of the luminosity distance in Eq.~\eqref{eq:dl_z0_tay}. Substituting these coefficients into the Padé expressions of Eq.~\eqref{eq:Pade_coeff}, we obtain
\begin{equation}
D_{\rm L}^{\mathrm{P(2,1)}}(z; z_0=0)=\frac{1}{2H_0} \left(\frac{6(1-q_0)z+(5-8q_0-3q_0^2+2j_0)z^2}{3(1-q_0)+(1-q_0-3q_0^2+j_0)z}\right),
\label{eq:dl_z0_Pade}
\end{equation}
which corresponds to the known expressions in ~\citet{Aviles:2014rma, Capozziello:2020ctn}.
On the other hand, for the expansion rate, we obtain
\begin{equation}
H^{\mathrm{P(2,1)}}(z; z_0=0)=\frac{ H_0 \left(3j_0^2 z^2+2j_0\,[z((q_0(q_0+7)+3)z+7q_0+6)+3]+9q_0^4 z^2-6q_0^2(z+1)^2+2q_0 s_0 z^2+2s_0 z(z+1)\right)}{ j_0((8q_0+6)z+6)+6q_0^3 z-6q_0^2(z+1)+2s_0 z}.
\end{equation}

\section{Numerical Implementation and Precision Tests} 
\label{section:B}

Building on the cosmographic framework outlined in the previous section, we have developed a dedicated Python module that fully implements the Padé-$(2,1)$ expansion around an arbitrary pivot redshift. Once a pivot redshift $z_0$ and the corresponding cosmographic parameters \{$H_{z_0}$, $q_{z_0}$, $j_{z_0}$, $s_{z_0}$\} are specified, the code makes direct use of the expressions derived in this work to compute the luminosity distance $D_L(z)$ via Eq.~\eqref{eq:ci_coefficients} and the Hubble parameter $H(z)$ via Eq.~\eqref{eq:Ci_coefficients}. From these quantities, all related distance measures are automatically obtained (such as the angular diameter distance $D_A(z)$ and the comoving distance $D_M(z)$) by enforcing the distance duality relation~\citep{Etherington1933,Ellis2007}.\footnote{For works spanning the past two decades testing the DDR against a multitude of astrophysical and cosmological datasets in both model-dependent and model-independent manners, we refer to ~\citet{Uzan:2004my,DeBernardis:2006ii,More:2008uq,Holanda:2010ay,Holanda:2010vb,Li:2011exa,Nair:2011dp,Liang:2011gm,Meng:2011nt,Holanda:2011hh,Khedekar:2011gf,Goncalves:2011ha,Lima:2011ye,Holanda:2012at,Cardone:2012vd,Yang:2013coa,Zhang:2014eux,Santos-da-Costa:2015kmv,Wu:2015ixa,Wu:2015ixa,Liao:2015uzb,Rana:2015feb,Ma:2016bjt,Holanda:2016msr,More:2016fca,Rana:2017sfr,Li:2017zrx,Lin:2018qal,Qi:2019spg,Holanda:2019vmh,Zhou:2020moc,Qin:2021jqy,Bora:2021cjl,Mukherjee:2021kcu,Liu:2021fka,Renzi:2021xii,Xu:2022zlm,Tonghua:2023hdz,Qi:2024acx,Yang:2024icv,Favale:2024sdq,Tang:2024zkc,Jesus:2024nrl,Qi:2024acx,Alfano:2025gie,Yang:2025qdg,Keil:2025ysb,Teixeira:2025czm,Ruchika:2025sbb,DeLeo:2025rhy}. For projected limits from future surveys, see, \textit{e.g.}, ~\citet{Cardone:2012vd,Yang:2017bkv,Fu:2019oll,Hogg:2020ktc,Renzi:2020bvl,Euclid:2020ojp,DeLeo:2025rhy}.} This framework offers several advantages. On the one hand, it allows us to validate the Padé expansion against exact predictions from numerical Boltzmann codes such as \texttt{CAMB}~\citep{Lewis:1999bs,Howlett:2012mh} and \texttt{CLASS}~\citep{Lesgourgues:2011re,Blas:2011rf,Lesgourgues:2011rg,Lesgourgues:2011rh} within specific cosmological models. For instance, by fixing the cosmographic parameters to their $\Lambda$CDM values inferred from the Planck best-fit cosmology, we can directly test how well the Padé-based reconstruction of $D_L(z)$ matches the exact model predictions. On the other hand, the module is fully interfaced with state-of-the-art cosmological samplers and will be released as part of a broader package for background analyses. Therefore, it provides a practical tool for confronting theoretical predictions with observational background data, enabling MCMC analyses directly in terms of cosmographic parameters.

\subsection{Precision Tests}

To assess the robustness of our method and to quantify the accuracy gained or lost when adopting different choices of the pivot redshift $z_0$, we perform a dedicated consistency test. The idea is to isolate the truncation error of the Padé-$(2,1)$ expansion and verify how it propagates across redshift when the series is expanded around different pivots. As a theoretical reference, we adopt a flat $\Lambda$CDM cosmology with all parameters fixed to their Planck 2018 best-fit values~\citep{Planck:2018vyg}. Using \texttt{CLASS}~\citep{Lesgourgues:2011re,Blas:2011rf,Lesgourgues:2011rg,Lesgourgues:2011rh}, we compute the exact background observables on a dense redshift grid up to $z=2.5$, which serves as the benchmark for our comparison.

\begin{figure}[htbp!]
    \centering
    \includegraphics[width=0.5\linewidth]{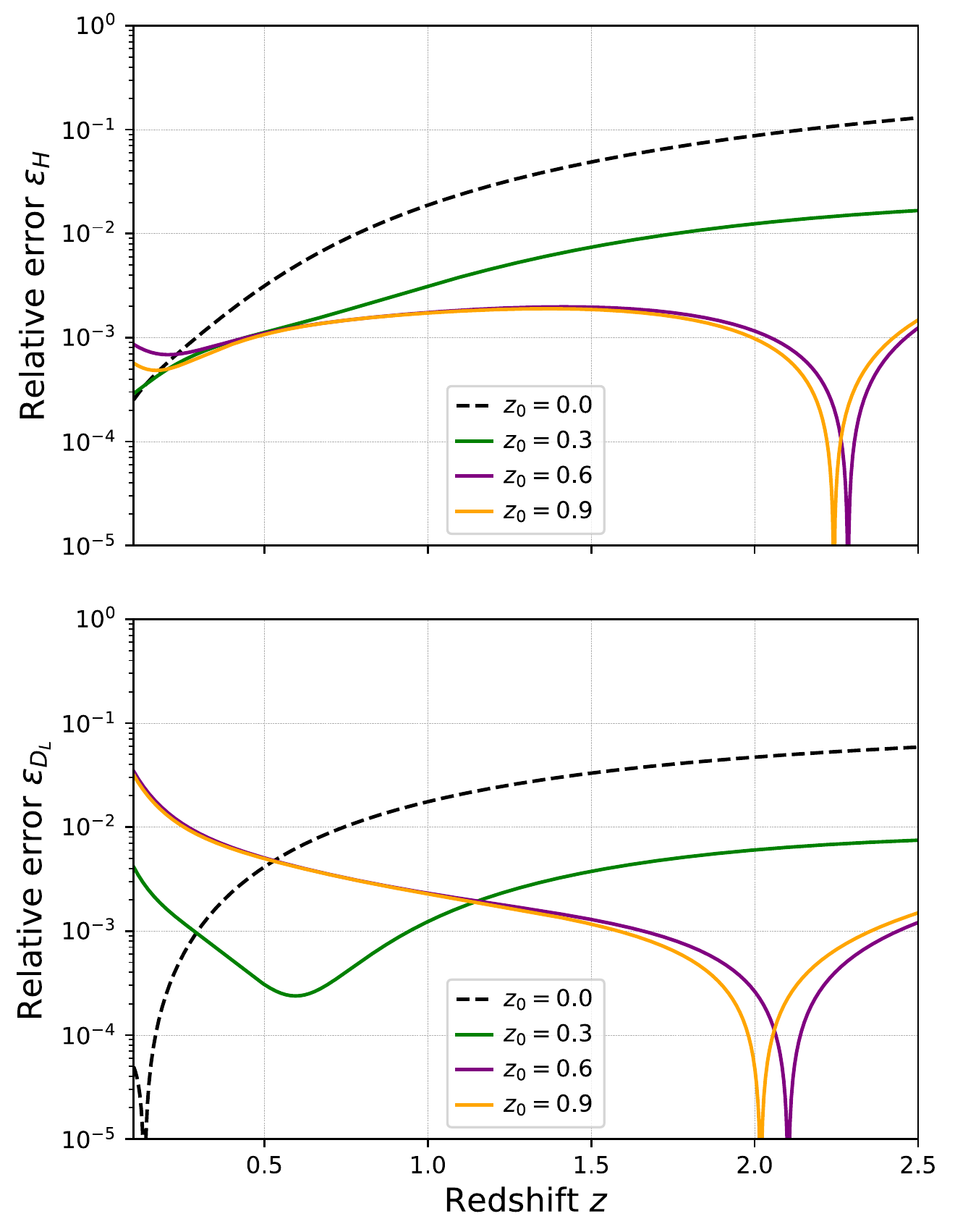}
    \caption{Comparison of the relative precision of the Padé expansion for $H(z)$ and $D_{\rm L}(z)$ at different redshifts $z$ (using exact $\Lambda$CDM predictions computed with CLASS as reference) for four representative choices of the pivot redshift $z_0$. The black dashed line shows the Padé expansion around $z_0=0$, corresponding to the standard case commonly discussed in the literature.}
\label{fig:S1}
\end{figure}

From this reference cosmology we derive the exact values of the cosmographic parameters $H(z)$, $q(z)$, $j(z)$, and $s(z)$ at any redshift $z$. These are obtained by combining the definitions in Eqs.~\eqref{eq:qz_def}, \eqref{eq:jz_def}, and \eqref{eq:sz_def} with the Hubble parameter in a flat $\Lambda$CDM model,
\begin{equation}
H_{z_0}=H_0 \sqrt{\Omega_{m}(1+z_0)^3+(1-\Omega_{m})}\,,
\label{eq:Hz0_LCDM}
\end{equation}
which leads to the following explicit expressions for the cosmographic parameters in $\Lambda$CDM, keeping $z_0$ as a free pivot:
\begin{equation}
q_{z_0}= \frac{\Omega_{m}(1+z_0)^3-2(1-\Omega_{m})}{2\,[\Omega_{m}(1+z_0)^3+(1-\Omega_{m})]}\,,
\label{eq:qz0_LCDM}
\end{equation}
\begin{equation}
j_{z_0}=1\,,
\label{eq:jz0_LCDM}
\end{equation}
\begin{equation}
s_{z_0}=1-\frac{9\,\Omega_{m}(1+z_0)^3}{2\,[\Omega_{m}(1+z_0)^3+(1-\Omega_{m})]}\,,
\label{eq:sz0_LCDM}
\end{equation}
where $H_0\equiv H(z=0)$ is the present-day Hubble constant and $\Omega_{m}$ denotes the present-day matter density parameter. For every chosen pivot $z_0$, we fix the cosmographic parameters to their exact values in $\Lambda$CDM at that pivot, i.e., $H_{z_0}$, $q_{z_0}$, $j_{z_0}\equiv 1$, and $s_{z_0}$. This ensures that the only difference between the Padé reconstruction and the exact $\Lambda$CDM observables is due to the truncation of the series, not to parameter estimation uncertainties. With these inputs, we build the Padé-$(2,1)$ expansion around $z_0$ and reconstruct the observables $H(z)$ and $D_{\rm L}(z)$ across the full redshift range $[0,2.5]$. We then compare the reconstructed quantities with the exact $\Lambda$CDM results from \texttt{CLASS}, defining the relative error as the absolute difference normalized to the exact value:
\begin{equation}
    \epsilon_X= \frac{X_{\Lambda \mathrm{CDM}}-X^{\mathrm{P}(2,1)}_{\Lambda \mathrm{CDM}}}{X_{\Lambda \mathrm{CDM}}} \,,
    \label{eq:rel_err}
\end{equation}
where $X$ indicates a cosmological function.

The results of this exercise are shown in Fig.~\ref{fig:S1}. The black dashed line corresponds to the standard case with $z_0=0$, which is commonly adopted in the literature. As expected, the relative error is very small close to the pivot but grows as one moves to higher redshift. In contrast, when adopting a pivot at $z_0=0.3$ (green line), the truncation error is slightly larger in the immediate vicinity of $z=0$, but is up to an order of magnitude smaller in the range $0.3\lesssim z\lesssim 2.5$, precisely where most of the data are concentrated. The same pattern holds for higher pivots: when setting $z_0=0.6$ (purple line) or $z_0=0.9$ (orange line), the error at low redshift $z\to0$ somewhat increases, but the expansion becomes up to two orders of magnitude more accurate at intermediate and high redshifts. The gain in accuracy is consistent for both $H(z)$ and $D_{\rm L}(z)$, and we have verified that it also applies to $D_{\rm V}(z)$ and to the ratio $D_{\rm M}(z)/D_{\rm H}(z)$.

Overall, this test illustrates a clear trade-off. Choosing a pivot at $z_0=0$ maximizes the accuracy at the present epoch but introduces increasingly large truncation errors at higher redshifts. Shifting the pivot to $z_0>0$ slightly reduces the accuracy at very low $z$ but substantially improves the reconstruction at the redshifts where most of the data lie.

We stress that the loss of precision at $z\simeq0$ is not problematic for our analysis, since when constraining cosmographic parameters at low $z_0$ we consistently use the Padé-$(2,1)$ expansion centered at low $z_0$. On the other hand, when extracting constraints at higher redshift, adopting an expansion centered closer to the redshift of interest yields clear gains in both accuracy and precision. This validates our multi-pivot strategy: by selecting different $z_0$ values, one can minimize the truncation error in the region most relevant for the data, achieving accurate reconstructions without relying on any dynamical assumptions. This demonstrates that the generalized Padé-$(2,1)$ expansion, when implemented around arbitrary pivots, provides a highly flexible and accurate tool for cosmographic reconstructions across the whole redshift range of interest.

We conclude this section by noting that the same precision tests remain valid when the cosmographic parameters are fixed to the values predicted by the best-fit Planck+DESI $w_0w_a$CDM model, where the DE equation of state follows the CPL parameterization $w(a)=w_0+w_a(1-a)$. In particular, performing the same derivation as in $\Lambda$CDM but using the Hubble parameter for a flat $w_0w_a$CDM model, we obtain the following relations for the cosmographic parameters:
\begin{equation}
q(z)=\frac{(1 - \Omega_{m})(1 + z)^{
    3 (w_0 + w_a)} (1 + z + 
      3 (w_0 + (w_0 + w_a) z))  + \Omega_{m} (1 + z)
   e^{\frac{3 w_a z}{1 + z}}}{2  (1 - \Omega_{m})(1 + z)^{
    1 + 3 w_0 + 3 w_a} +
   2 \Omega_{m} (1 + z) e^{\frac{3 w_a z}{1 + z}}} \,,
   \label{eq:qz_CPL}
\end{equation}
\begin{equation}
\begin{aligned}
j(z) =& 
\frac{(1 - \Omega_{m}) (1 + z)^{3(w_0 + w_a)}
   \bigl[2 + 3w_a
   + 4(1 + 3w_a)z + (2 + 9w_a(1 + w_a))z^2 
         + 9w_0^2(1 + z)^2 }
     {2(1 + z)^2\bigl[(1 - \Omega_{m})(1 + z)^{3(w_0 + w_a)}
                     + \Omega_{m} e^{\frac{3w_a z}{1 + z}}\bigr]} \\
&\frac{+ 9w_0(1 + z)(1 + z + 2w_a z)\bigr]+2\Omega_{m}(1 + z)^2e^{\frac{3w_a z}{1 + z}}}
     {2(1 + z)^2\bigl[(1 - \Omega_{m})(1 + z)^{3(w_0 + w_a)}
                     + \Omega_{m} e^{\frac{3w_a z}{1 + z}}\bigr]} \,,
\end{aligned}
   \label{eq:jz_CPL}
\end{equation}
where $w_0$ and $w_a$ are the two parameters governing the evolution of dark energy. We have explicitly verified that the same precision tests performed for $\Lambda$CDM remain valid in the CPL framework when the cosmographic parameters are fixed to their best-fit values obtained under CPL priors, exactly as we did for $\Lambda$CDM. In addition, these are the expressions used in the main text to compare the CPL predictions with the cosmographic reconstructions.

\subsection{Markov Chains Monte Carlo Analyses}

The results presented in the main text are obtained from Markov Chain Monte Carlo (MCMC) analyses carried out for different fixed values of the pivot redshift $z_0$. In particular, we explore a uniform grid in the range $z_0 \in [0,1]$ with step size $\Delta z_0=0.1$. To compare the theoretical predictions with data, we employ the following background-level datasets:
\begin{itemize}
    \item \textbf{Type Ia Supernovae} (SNIa): Distance moduli measurements relating the theoretical luminosity distance with the magnitude–redshift relation $\mu(z)=5\log(D_{\rm L}(z))+25$. We use the PantheonPlus (PP)~\citep{Brout:2022vxf} catalog and the 5-year dataset of the Dark Energy Survey (DESy5)~\citep{DES:2024jxu}. We consider these two catalogs because the DESI Collaboration~\citep{DESI:2025zgx} showed that they represent, respectively, the combinations with the minimum and maximum preference for evolving Dark Energy scenarios. 
    \item \textbf{Cosmic Chronometers} (CC): Measurements of the Hubble parameter $H(z)$~\citep{Jimenez:2001gg, Borghi:2021rft}, which can be directly compared against the theoretical Padé expansion of the same function. We selected 15 data points from~\citep{Moresco:2012jh, Moresco:2015cya, Moresco:2016mzx}, since for this subset full estimates of the covariance matrix’s non-diagonal terms and systematic contributions are available, as outlined in ~\citet{Moresco:2018xdr, Moresco:2020fbm}. We exclude some earlier measurements due to concerns expressed in ~\citet{Kjerrgren:2021zuo}, which do not apply to our selected data. This dataset is referred to as “CC”.
    \item \textbf{Baryon Acoustic Oscillations} (BAO): Data from the Dark Energy Spectroscopic Instrument (DESI) second data release~\citep{DESI:2025zgx}. These measurements include the transverse comoving distance $D_\mathrm M(z)$, the Hubble horizon $D_\mathrm H(z)$, and the angle-averaged distance $D_\mathrm V(z)$, all normalized to the sound horizon at the drag epoch $r_\mathrm d$. Observational data are compared with the respective theoretical functions obtained from the Padé expansions of the expansion rate $H(z)$ and luminosity distance $D_\mathrm L(z)$, using the relations $D_{\rm L}(z)=(1+z)D_\mathrm M(z)$, $D_\mathrm H(z)=c/H(z)$, and $D_\mathrm V(z)=(z D_\mathrm M(z)^2 D_\mathrm H(z))^{1/3}$. Since we are performing a purely late-time background analysis, the sound horizon $r_d$ is calibrated using Planck $\Lambda$CDM cosmology~\citep{Planck:2018vyg}, with a Gaussian prior $r_\mathrm d=(147.09\pm0.26)$ Mpc. This choice is both natural and robust: models of dynamical Dark Energy cannot alter the physics before recombination that determines $r_d$. Moreover, Padé and Taylor expansions are only accurate at low redshift, so we cannot extrapolate them up to $z\sim1100$ and include the CMB as an effective BAO point, as is sometimes done in the literature. Instead, calibrating BAO with the Planck-inferred $r_d$ provides a consistent way to anchor our low-redshift analysis. We denote this combined dataset as “DESI+$r_d$”. For completeness, in \hyperref[section:D]{Appendix D} we explore alternative BAO (and SNIa) calibrations motivated by the Hubble tension, showing that they only rescale $H(z)$ and do not affect the cosmographic conclusions.
\end{itemize}
The MCMC analysis is carried out with our dedicated numerical package, fully interfaced with the \texttt{Cobaya} software~\citep{Torrado:2020dgo}, where a Markov Chain Monte Carlo (MCMC) sampler is implemented to compare the cosmological functions derived from the Padé expansion with the observational datasets. Uniform priors are adopted for the free cosmographic parameters: $H_{z_0}\in\mathcal{U}[40,140]$ km/s/Mpc, $q_{z_0}\in\mathcal{U}[-2, 2]$, $j_{z_0}\in\mathcal{U}[-5, 5]$, and $s_{z_0}\in\mathcal{U}[-5, 5]$. Convergence of the chains is assessed through the Gelman–Rubin criterion, requiring $\mathrm{R}-1<0.01$~\citep{Gelman:1992zz}.

\section{Supplementary Results and Discussions} 
\label{section:C}

\begin{table*}[htpb!]
\centering
\caption{Constraints at 68\% CL on the expansion rate $H(z_0)$ (in km/s/Mpc) and on the cosmographic parameters $q(z_0)$, $j(z_0)$, and $s(z_0)$ at different fixed pivot redshifts $z_0$, for various combinations of datasets. For each dataset combination, results are reported both without including cosmic chronometers and including them, the latter shown in round brackets.}
\vspace{0.1cm}
\renewcommand{\arraystretch}{0.9}
\resizebox{\textwidth}{!}{
\begin{tabular}{l  l c c c c c c c c c}
\hline\hline
\\
\textbf{Redshift}  &  \textbf{Dataset}            &  \boldmath{$H(z_0)$}         &  \boldmath{$q(z_0)$}                &  \boldmath{$j(z_0)$}               &  \boldmath{$s(z_0)$}             \\
\\
\hline
&&&&&\\

$z_0=0.0$  &  DESI+$r_d$\,(+CC)  &  $67.49\pm0.56\;(66.4\pm1.0)$  &  $-0.427^{+0.058}_{-0.051}\;(-0.336^{+0.099}_{-0.078})$  &  $0.78^{+0.23}_{-0.30}\;(0.50^{+0.23}_{-0.36})$  &  $-0.26^{+0.26}_{-0.50}\;(-0.37^{+0.29}_{-0.57})$\\
           &  PP\,(+CC)          & Unconstrained $(66.8\pm 4.3)$  &  $-0.504\pm 0.089\;(-0.507\pm0.071)$  &  $1.33^{+0.69}_{-0.81}\;(1.35\pm0.53)$  &  Unconstrained $(1.8^{+2.5}_{-1.5})$\\
           &  DESy5\,(+CC)       & Unconstrained $(66.8\pm 4.1)$  &  $-0.405\pm 0.092\;(-0.426\pm0.086)$  &  $0.80^{+0.63}_{-0.74}\;(0.96\pm0.59)$  &  Unconstrained $(1.4^{+2.8}_{-1.6})$\\
           &  DESI+$r_d$+PP\,(+CC)   &  $67.51\pm0.58\;(67.49\pm0.56)$  &  $-0.426^{+0.063}_{-0.054}\;(-0.427^{+0.058}_{-0.051})$  &  $0.77^{+0.25}_{-0.34}\;(0.78^{+0.23}_{-0.30})$  &  $-0.23^{+0.28}_{-0.62}\;(-0.26^{+0.26}_{-0.50})$\\
           &  DESI+$r_d$+DESy5\,(+CC) &  $66.89\pm0.53\;(66.90\pm0.53)$ & $-0.362\pm0.041\;(-0.366^{+0.048}_{-0.042})$ & $0.53^{+0.15}_{-0.19}\;(0.55^{+0.16}_{-0.20})$ & $-0.491^{+0.068}_{-0.11}\;(-0.475^{+0.099}_{-0.16})$\\[1ex]

\hline \\[1ex]
$z_0=0.1$  &  DESI+$r_d$\,(+CC) & $71.61\pm0.45\;(71.55\pm0.87)$ & $-0.379\pm0.039\;(-0.38^{+0.14}_{-0.12})$ & $0.87^{+0.23}_{-0.27}\;(0.95^{+0.53}_{-0.75})$ & $-0.42^{+0.28}_{-0.48}\;(<-0.0559)$\\
           &  PP\,(+CC)          & Unconstrained $(70.6\pm4.3)$ & $-0.392\pm0.047\;(-0.401\pm0.044)$ & $1.08^{+0.53}_{-0.59}\;(1.23\pm0.46)$ & Unconstrained $(1.3\pm1.9)$\\
           &  DESy5\,(+CC)       & Unconstrained $(70.8\pm4.5)$ & $-0.325\pm0.048\;(-0.338\pm0.049)$ & $0.65^{+0.48}_{-0.54}\;(0.84\pm0.49)$ & Unconstrained $(0.97^{+2.5}_{-2.0})$\\
           &  DESI+$r_d$+PP\,(+CC)   & $71.62\pm0.46\;(71.61\pm0.45)$ & $-0.374\pm0.041\;(-0.379\pm0.039)$ & $0.84^{+0.23}_{-0.28}\;(0.87^{+0.23}_{-0.27})$ & $-0.45^{+0.24}_{-0.47}\;(-0.42^{+0.28}_{-0.48})$\\
           &  DESI+$r_d$+DESy5\,(+CC) & $71.34\pm0.44\;(71.31\pm0.43)$ & $-0.324\pm0.037\;(-0.325\pm0.035)$ & $0.62^{+0.18}_{-0.22}\;(0.63^{+0.17}_{-0.20})$ & $-0.65^{+0.13}_{-0.20}\;(-0.671^{+0.094}_{-0.15})$\\[1ex]

\hline \\[1ex]
$z_0=0.2$  &  DESI+$r_d$\,(+CC) & $75.83\pm0.41\;(75.91\pm0.46)$ & $-0.319\pm0.032\;(-0.333\pm0.087)$ & $0.96\pm0.23\;(1.06^{+0.49}_{-0.63})$ & $-0.58^{+0.33}_{-0.54}\;(-0.08^{+0.53}_{-1.7})$\\
           &  PP\,(+CC)          & Unconstrained $(74.5\pm4.6)$ & $-0.325\pm0.032\;(-0.325\pm0.031)$ & $1.31\pm0.45\;(1.41\pm0.40)$ & Unconstrained $(1.0^{+2.7}_{-2.1})$\\
           &  DESy5\,(+CC)       & Unconstrained $(74.9\pm4.5)$ & $-0.276\pm0.032\;(-0.280\pm0.032)$ & $0.95^{+0.60}_{-0.67}\;(1.14\pm0.55)$ & Unconstrained $(1.0^{+2.9}_{-2.0})$\\
           &  DESI+$r_d$+PP\,(+CC)   & $75.86\pm0.42\;(75.83\pm0.41)$ & $-0.322\pm0.031\;(-0.319\pm0.032)$ & $0.98^{+0.23}_{-0.26}\;(0.96\pm0.23)$ & $-0.55^{+0.35}_{-0.59}\;(-0.58^{+0.33}_{-0.54})$\\
           &  DESI+$r_d$+DESy5\,(+CC) & $75.81\pm0.39\;(75.79\pm0.41)$ & $-0.274\pm0.028\;(-0.276\pm0.029)$ & $0.69^{+0.17}_{-0.20}\;(0.72^{+0.19}_{-0.22})$ & $-0.89^{+0.11}_{-0.17}\;(-0.85^{+0.15}_{-0.24})$\\[1ex]

\hline \\[1ex]
$z_0=0.3$  &  DESI+$r_d$\,(+CC) & $80.71\pm0.35\;(80.42\pm0.46)$ & $-0.235\pm0.022\;(-0.249\pm0.057)$ & $0.73\pm0.11\;(0.97^{+0.40}_{-0.49})$ & $-1.186^{+0.077}_{-0.098}\;(-0.62^{+0.51}_{-1.2})$\\
           &  PP\,(+CC)          & Unconstrained $(80.7\pm4.7)$ & $-0.260\pm0.032\;(-0.255\pm0.031)$ & $0.66\pm0.14\;(0.69\pm0.14)$ &  Unconstrained $(0.8\pm2.0)$\\
           &  DESy5\,(+CC)       &  Unconstrained $(79.9\pm4.8)$ & $-0.232\pm0.031\;(-0.233\pm0.030)$ & $1.02\pm0.31\;(1.10\pm0.29)$ &  Unconstrained $(0.6\pm2.3)$\\
           &  DESI+$r_d$+PP\,(+CC)   & $80.73\pm0.35\;(80.71\pm0.35)$ & $-0.236\pm0.022\;(-0.235\pm0.022)$ & $0.72\pm0.11\;(0.73\pm0.11)$ & $-1.187^{+0.077}_{-0.096}\;(-1.186^{+0.077}_{-0.098})$\\
           &  DESI+$r_d$+DESy5\,(+CC) & $80.41\pm0.41\;(80.38\pm0.41)$ & $-0.234\pm0.027\;(-0.238\pm0.026)$ & $0.88^{+0.22}_{-0.25}\;(0.92\pm0.22)$ & $-0.96^{+0.25}_{-0.45}\;(-0.92^{+0.28}_{-0.47})$\\[1ex]
\hline \\[1ex]
$z_0=0.4$  &  DESI+$r_d$\,(+CC) & $85.85\pm0.31\;(85.40\pm0.57)$ & $-0.165\pm0.018\;(-0.170\pm0.038)$ & $0.591\pm0.042\;(0.87^{+0.34}_{-0.41})$ & $-1.444\pm0.064\;(-1.06^{+0.44}_{-0.79})$\\
           &  PP\,(+CC)          &  Unconstrained $(85.1\pm5.2)$ & $-0.209\pm0.034\;(-0.200\pm0.033)$ & $0.561\pm0.050\;(0.567\pm0.049)$ &  Unconstrained $(0.8^{+2.7}_{-2.3})$\\
           &  DESy5\,(+CC)       &  Unconstrained $(85.4\pm5.2)$ & $-0.180\pm0.034\;(-0.177\pm0.035)$ & $0.690\pm0.098\;(0.69\pm0.10)$ &  Unconstrained $(0.8\pm2.2)$\\
           &  DESI+$r_d$+PP\,(+CC)   & $85.84\pm0.31\;(85.85\pm0.31)$ & $-0.165\pm0.018\;(-0.165\pm0.018)$ & $0.590\pm0.043\;(0.591\pm0.042)$ & $-1.444\pm0.067\;(-1.444\pm0.064)$\\
           &  DESI+$r_d$+DESy5\,(+CC) & $85.65\pm0.33\;(85.64\pm0.33)$ & $-0.166\pm0.020\;(-0.167\pm0.020)$ & $0.694\pm0.091\;(0.698\pm0.090)$ & $-1.458^{+0.090}_{-0.10}\;(-1.463^{+0.087}_{-0.10})$\\[1ex]

\hline \\[1ex]
$z_0=0.5$  &  DESI+$r_d$\,(+CC) & $91.10\pm0.31\;(90.70\pm0.60)$ & $-0.110\pm0.018\;(-0.108\pm0.026)$ & $0.611\pm0.026\;(0.84^{+0.30}_{-0.35})$ & $-1.703\pm0.082\;(-1.41^{+0.37}_{-0.60})$\\
           &  PP\,(+CC)          & Unconstrained $\;(90.2\pm5.2)$ & $-0.156\pm0.038\;(-0.147\pm0.038)$ & $0.571\pm0.036\;(0.577\pm0.038)$ & Unconstrained $(0.5^{+2.7}_{-2.4})$\\
           &  DESy5\,(+CC)       & Unconstrained $(90.3\pm5.2)$ & $-0.132\pm0.037\;(-0.122\pm0.036)$ & $0.642\pm0.041\;(0.646\pm0.042)$ & Unconstrained $(0.7\pm2.3)$\\
           &  DESI+$r_d$+PP\,(+CC)   & $91.08\pm0.31\;(91.10\pm0.31)$ & $-0.112\pm0.017\;(-0.110\pm0.018)$ & $0.608\pm0.024\;(0.611\pm0.026)$ & $-1.704\pm0.078\;(-1.703\pm0.082)$\\
           &  DESI+$r_d$+DESy5\,(+CC) & $91.01\pm0.31\;(90.99\pm0.31)$ & $-0.107\pm0.018\;(-0.108\pm0.019)$ & $0.652\pm0.038\;(0.653\pm0.038)$ & $-1.717^{+0.097}_{-0.11}\;(-1.724^{+0.095}_{-0.11})$\\[1ex]

\hline \\[1ex]
$z_0=0.6$  &  DESI+$r_d$\,(+CC) & $96.54\pm0.33\;(96.25\pm0.64)$ & $-0.060\pm0.018\;(-0.050\pm0.022)$ & $0.654\pm0.024\;(0.85^{+0.29}_{-0.32})$ & $-1.99^{+0.11}_{-0.13}\;(-1.62^{+0.46}_{-0.72})$\\
           &  PP\,(+CC)          &  Unconstrained $(95.1\pm5.7)$ & $-0.103\pm0.040\;(-0.089\pm0.040)$ & $0.607^{+0.043}_{-0.049}\;(0.623^{+0.046}_{-0.051})$ &  Unconstrained $(-0.2\pm2.1)$\\
           &  DESy5\,(+CC)       &  Unconstrained $(95.4\pm5.5)$ & $-0.075\pm0.043\;(-0.067\pm0.041)$ & $0.664\pm0.044\;(0.672\pm0.044)$ &  Unconstrained $(0.0\pm2.2)$\\
           &  DESI+$r_d$+PP\,(+CC)   & $96.54\pm0.32\;(96.54\pm0.33)$ & $-0.062\pm0.018\;(-0.060\pm0.018)$ & $0.652\pm0.025\;(0.654\pm0.024)$ & $-1.99^{+0.11}_{-0.13}\;(-1.99^{+0.11}_{-0.13})$\\
           &  DESI+$r_d$+DESy5\,(+CC) & $96.52\pm0.34\;(96.54\pm0.34)$ & $-0.057\pm0.019\;(-0.056\pm0.019)$ & $0.680\pm0.026\;(0.682\pm0.027)$ & $-1.99^{+0.12}_{-0.15}\;(-1.98^{+0.13}_{-0.16})$\\[1ex]

\hline \\[1ex]
$z_0=0.7$  &  DESI+$r_d$\,(+CC) & $102.30\pm0.39\;(102.08\pm0.64)$ & $-0.008\pm0.020\;(0.001\pm0.023)$ & $0.712\pm0.028\;(0.88\pm0.27)$ & $-2.25^{+0.18}_{-0.23}\;(-1.85^{+0.50}_{-0.79})$\\
           &  PP\,(+CC)          &  Unconstrained $(99.98^{+7.5}_{-6.7})$ & $-0.039\pm0.045\;(-0.026\pm0.045)$ & $0.673\pm0.059\;(0.690\pm0.061)$ &  Unconstrained $(-0.7^{+2.0}_{-2.2})$\\
           &  DESy5\,(+CC)       &  Unconstrained $(100.8\pm6.0)$ & $-0.019\pm0.046\;(-0.009\pm0.046)$ & $0.717^{+0.057}_{-0.065}\;(0.729\pm0.059)$ &  Unconstrained $(-0.9^{+2.0}_{-2.4})$\\
           &  DESI+$r_d$+PP\,(+CC)   & $102.28\pm0.38\;(102.30\pm0.39)$ & $-0.0097\pm0.019\;(-0.008\pm0.020)$ & $0.710\pm0.028\;(0.712\pm0.028)$ & $-2.26^{+0.17}_{-0.22}\;(-2.25^{+0.18}_{-0.23})$\\
           &  DESI+$r_d$+DESy5\,(+CC) & $102.27\pm0.39\;(102.29\pm0.39)$ & $-0.007\pm0.019\;(-0.005\pm0.019)$ & $0.729\pm0.027\;(0.732\pm0.028)$ & $-2.26^{+0.18}_{-0.23}\;(-2.23^{+0.19}_{-0.25})$\\[1ex]

\hline \\[1ex]
$z_0=0.8$  &  DESI+$r_d$\,(+CC) & $108.33\pm0.47\;(108.23\pm0.67)$ & $0.043\pm0.021\;(0.049\pm0.024)$ & $0.777\pm0.033\;(0.87\pm0.25)$ & $-2.46^{+0.30}_{-0.38}\;(-2.12^{+0.54}_{-0.77})$\\
           &  PP\,(+CC)          &  Unconstrained $(106.9\pm6.3)$ & $0.029\pm0.053\;(0.042\pm0.051)$ & $0.758\pm0.081\;(0.779\pm0.081)$ &  Unconstrained $(-1.5^{+1.7}_{-2.3})$\\
           &  DESy5\,(+CC)       &  Unconstrained $(107.4\pm6.3)$ & $0.048\pm0.051\;(0.057\pm0.049)$ & $0.800^{+0.077}_{-0.087}\;(0.812\pm0.076)$ &  Unconstrained $(-1.6^{+1.6}_{-2.4})$\\
           &  DESI+$r_d$+PP\,(+CC)   & $108.33\pm0.48\;(108.33\pm0.47)$ & $0.041\pm0.021\;(0.043\pm0.021)$ & $0.774\pm0.033\;(0.777\pm0.033)$ & $-2.49^{+0.29}_{-0.39}\;(-2.46^{+0.30}_{-0.38})$\\
           &  DESI+$r_d$+DESy5\,(+CC) & $108.36\pm0.48\;(108.35\pm0.48)$ & $0.046\pm0.020\;(0.046\pm0.021)$ & $0.793\pm0.032\;(0.793\pm0.032)$ & $-2.43^{+0.33}_{-0.45}\;(-2.44^{+0.32}_{-0.44})$\\[1ex]

\hline \\[1ex]
$z_0=0.9$  &  DESI+$r_d$\,(+CC) & $114.78\pm0.58\;(114.61\pm0.75)$ & $0.098\pm0.023\;(0.098\pm0.025)$ & $0.856\pm0.039\;(0.91^{+0.25}_{-0.22})$ & $-2.49^{+0.56}_{-0.78}\;(-2.29^{+0.67}_{-0.90})$\\
           &  PP\,(+CC)          &  Unconstrained $(113.8\pm7.3)$ & $0.118\pm0.059\;(0.123\pm0.060)$ & $0.90^{+0.11}_{-0.12}\;(0.91^{+0.11}_{-0.12})$ &  Unconstrained $(-2.1^{+1.3}_{-2.2})$\\
           &  DESy5\,(+CC)       &  Unconstrained $(113.9\pm6.9)$ & $0.123\pm0.057\;(0.133\pm0.058)$ & $0.91\pm0.10\;(0.93^{+0.10}_{-0.12})$ &  Unconstrained $(-2.18^{+0.92}_{-2.6})$\\
           &  DESI+$r_d$+PP\,(+CC)   & $114.70\pm0.59\;(114.78\pm0.58)$ & $0.095\pm0.022\;(0.098\pm0.023)$ & $0.851\pm0.038\;(0.856\pm0.039)$ & $-2.58^{+0.53}_{-0.73}\;(-2.49^{+0.56}_{-0.78})$\\
           &  DESI+$r_d$+DESy5\,(+CC) & $114.78\pm0.58\;(114.78\pm0.59)$ & $0.099\pm0.022\;(0.099\pm0.022)$ & $0.867\pm0.038\;(0.867\pm0.038)$ & $-2.47^{+0.58}_{-0.81}\;(-2.47^{+0.58}_{-0.79})$\\[1ex]

\hline \\[1ex]
$z_0=1.0$  &  DESI+$r_d$\,(+CC) & $121.32^{+0.84}_{-0.70}\;(121.12\pm0.85)$ & $0.144^{+0.030}_{-0.025}\;(0.141\pm0.024)$ & $0.926^{+0.054}_{-0.046}\;(0.96\pm0.20)$ & $-2.64^{+0.87}_{-1.0}\;(-2.67^{+0.71}_{-0.88})$\\
           &  PP\,(+CC)          &  Unconstrained $(122.0^{+8.3}_{-7.5})$ & $0.216\pm0.071\;(0.218\pm0.070)$ & $1.07^{+0.15}_{-0.16}\;(1.08^{+0.14}_{-0.16})$ &  Unconstrained $(-2.6^{+1.0}_{-1.9})$\\
           &  DESy5\,(+CC)       &  Unconstrained $(121.8^{+8.3}_{-7.3})$ & $0.220\pm0.069\;(0.220\pm0.066)$ & $1.09^{+0.14}_{-0.16}\;(1.09^{+0.13}_{-0.15})$ &  Unconstrained $(-2.7^{+1.0}_{-1.8})$\\
           &  DESI+$r_d$+PP\,(+CC)   & $121.48\pm0.66\;(121.32^{+0.84}_{-0.70})$ & $0.151\pm0.023\;(0.144^{+0.030}_{-0.025})$ & $0.939\pm0.042\;(0.926^{+0.054}_{-0.046})$ & $-2.37^{+0.87}_{-1.1}\;(-2.64^{+0.87}_{-1.0})$\\
           &  DESI+$r_d$+DESy5\,(+CC) & $121.44\pm0.66\;(121.43\pm0.63)$ & $0.151\pm0.022\;(0.149\pm0.021)$ & $0.947\pm0.041\;(0.945\pm0.039)$ & $-2.38^{+0.85}_{-1.1}\;(-2.50^{+0.78}_{-0.90})$\\[1ex]

\hline
\bottomrule
\end{tabular}}
\label{tab:results}
\end{table*}

\begin{figure}[htpb!]
    \centering
    \includegraphics[width=\linewidth]{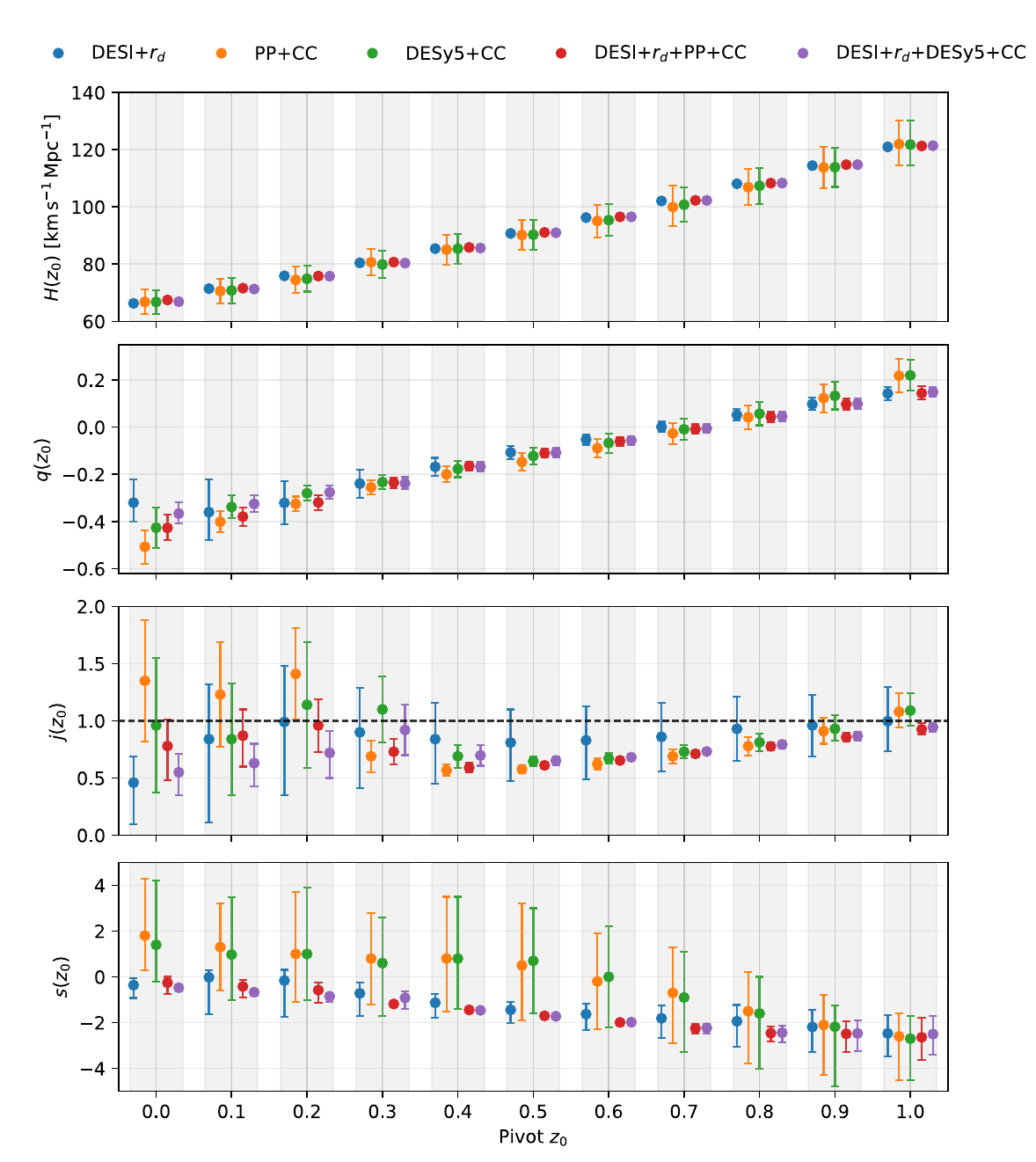}
    \caption{Whisker plot summarizing the 68\% CL constraints on the cosmographic parameters $H(z_0)$, $q(z_0)$, $j(z_0)$, and $s(z_0)$ at different pivot redshifts $z_0\in[0,1]$, for all dataset combinations considered in this work. For clarity, the constraints are slightly displaced horizontally around each pivot.}
    \label{fig:S2}
\end{figure}

\begin{figure}[htpb!]
    \centering
    \includegraphics[width=0.9\linewidth]{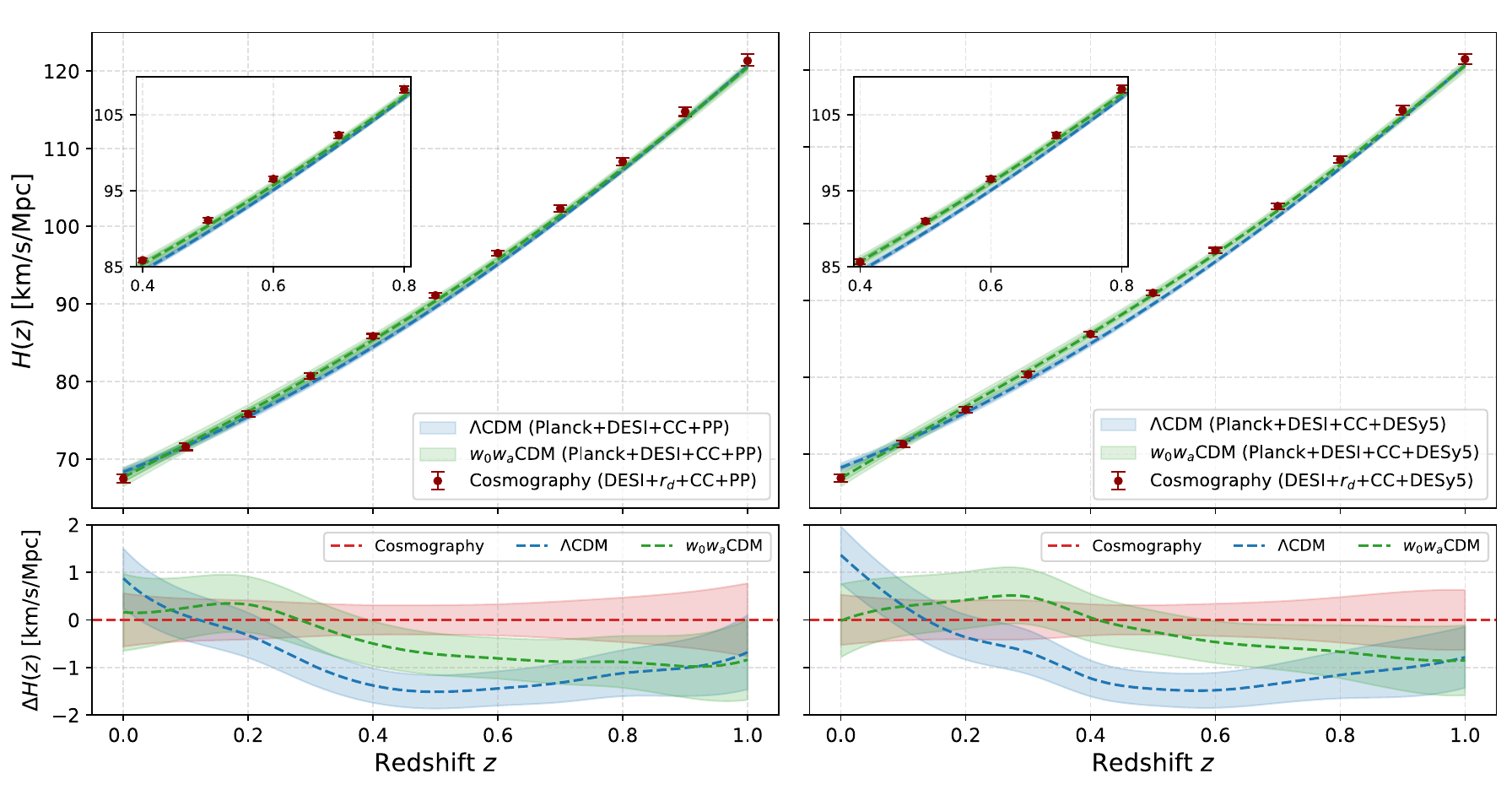}
    \caption{Cosmographic constraints on the Hubble parameter $H(z)$ at different redshift bins with 68\% CL uncertainties (red points). The blue band shows the $\Lambda$CDM predictions, while the green band corresponds to those from a CPL model of dynamical DE. The left panel shows results obtained by combining Planck-calibrated DESI BAO, CC, and PP SNIa, while the right panel shows results from the same data combination but with DESy5 SNIa instead of PP. The lower panels display the residuals, showing that the cosmographic reconstruction consistently prefers values of $H(z)$ about $2$ km/s/Mpc higher than $\Lambda$CDM, in better agreement with CPL expectations.}
    \label{fig:S3}
\end{figure}

In this section we present the results of the comprehensive analysis carried out for the different data combinations listed previously. Tab.~\ref{tab:results} reports the numerical constraints on the expansion rate $H(z_0)$ and on the cosmographic parameters $q(z_0)$, $j(z_0)$, and $s(z_0)$, derived at various pivot redshifts and for all dataset combinations considered. For every dataset combination in Tab.~\ref{tab:results}, we repeat the full cosmographic analysis both including and excluding CC. These measurements are known to rely more strongly on astrophysical modeling than BAO or SNIa, since they are derived from differential age estimates of passively evolving galaxies. In combinations that include DESI BAO calibrated with the Planck prior on $r_d$ together with SNIa (i.e., the “most informative” combinations discussed in the main text), the contribution of CC is naturally limited. DESI+$r_d$ constrain the expansion rate very tightly, and both PP and DESy5 SNIa provide high-precision distance-modulus measurements, leaving CC with comparatively little statistical weight. Their leverage on $H(z_0)$, $q(z_0)$, $j(z_0)$, and $s(z_0)$ is therefore small, as reflected in Tab.~\ref{tab:results}: for these combinations, the constraints obtained with and without CC are nearly identical at all pivot redshifts. This confirms that the BAO+$r_d$+SNIa results shown in the main text are not significantly affected by the inclusion of CC. A similar argument applies when comparing BAO+$r_d$ alone with BAO+$r_d$+CC: CC do not modify the qualitative conclusions because the expansion history is already tightly constrained by calibrated BAO. By contrast, the role of CC becomes important in SNIa-only combinations. SNIa alone cannot constrain $H(z_0)$ because of the well-known degeneracy with the absolute magnitude $M$. This degeneracy directly propagates into the snap parameter $s(z_0)$, which depends sensitively on the absolute expansion scale. This is explicitly reflected in the fact that the coefficients $c_i$ in Eq.\eqref{eq:ci_coefficients} for the expansion of $D_L(z)$ in Eq.\eqref{eq:Tayalor_ci_dl} depend only on the quantities $A$ and $B$ defined in Eqs.\eqref{eq:A_def} and \eqref{eq:B_def}, which in turn involve only $q(z)$ and $j(z)$. Conversely, the coefficient $c_3$ in Eq.\eqref{eq:Ci_coefficients} for the expansion of $H(z)$ in Eq.\eqref{eq:HTay_Ci} also depends on the quantity $C$ defined in Eq.\eqref{eq:C_def}, which depends on $s(z)$. As a result, in SNIa-only analyses both $H(z_0)$ and $s(z_0)$ remain unconstrained when CC are excluded, as shown in Tab.~\ref{tab:results}. Adding CC provides an absolute measurement of the Hubble rate at multiple redshifts, breaking the $H$–$M$ degeneracy and enabling meaningful (though still relatively broad) constraints on $H(z_0)$ and $s(z_0)$.

Having clarified the role of CC (i.e., the fact that they do not significantly alter the constraints in combinations where parameters are already tightly constrained by DESI+$r_d$ or DESI+$r_d$+SNIa, while being essential for constraining $H(z_0)$ and $s(z_0)$ in SNIa-only analyses) we now focus on the results obtained from datasets that include the additional information provided by CC. A compact visual representation of these results is provided in the whisker plot in Fig.~\ref{fig:S2}. While the main highlights and implications of these findings have been discussed in the main text, here we present complementary consistency checks and extended discussions. These results, on the one hand, reinforce the robustness of our analysis and, on the other, provide additional perspective on the behavior of the parameters across datasets.

The global picture that emerges is one of remarkable consistency across the results obtained with different data combinations. The whisker plot shows that, at each pivot, the constraints from independent probes are always compatible within $1\sigma$, with only very rare cases approaching $2\sigma$. This cross-consistency is an important outcome of our analysis, since it demonstrates that the cosmographic reconstruction delivers stable results across datasets of very different nature, from Planck-calibrated BAO to SNIa combined with cosmic chronometers.

Focusing first on the deceleration parameter, the results follow the expected trend: $q(z_0)$ is negative at low redshift, approaches zero around $z_0\simeq0.7$, and becomes positive afterwards. Interestingly, at very low redshift DESI+$r_d$ prefers less negative values of $q_0$ than the SNIa-based combinations, which point to a more strongly accelerated expansion. For instance, at $z_0=0$ we find $q=-0.320^{+0.10}_{-0.079}$ from DESI+$r_d$, while PP+CC yields $q=-0.507\pm0.071$. Although these differences remain compatible within $\sim2\sigma$, it is worth noting that larger values of $q_0$ are naturally expected in dynamical dark energy scenarios. Therefore, this suggests that DESI BAO data may strengthen the shift towards the predictions of dynamical dark energy compared to SNIa, while still remaining consistent overall.

The behavior of the jerk parameter is more subtle. At very low pivot redshifts, both BAO and SNIa provide only weak constraints on $j(z_0)$. This is expected: at low redshift the contribution of the jerk parameter is subdominant compared to the deceleration parameter, and its effect can only be constrained through extrapolation to higher redshift. In addition, fixing the pivot redshift to $z_0\sim0$, the inferred constraints are broadened by a strong correlation between $j(z_0)$ and $q(z_0)$. From Fig.~\ref{fig:S2}, we see that at $z_0\simeq0$ DESI+$r_d$ alone even favors values below unity, though with large uncertainties. Conversely, when the expansion is pivoted at higher redshifts, the sensitivity to $j$ increases significantly. This is due to three effects: first, the impact of $j$ on the cosmographic expansion naturally grows with redshift; second, expanding around higher pivots reduces correlations among parameters, allowing a cleaner determination of their contributions, as described in the main text; and third, the large number of SNIa in the range $0.3\lesssim z \lesssim 0.8$ makes their contribution especially powerful at these redshifts. In this regime SNIa deliver precise constraints on $j$ and reveal the most pronounced deviations from the $\Lambda$CDM prediction $j=1$. For example, DESI+$r_d$+PP+CC gives $j(0.4)=0.591\pm0.042$ and $j(0.8)=0.777\pm0.033$. Notably, SNIa in some cases even outperform BAO in constraining $j$, since BAO are intrinsically more sensitive to the snap parameter, as discussed below. Overall, the complementarity between probes is very clear in the combined analyses, where the inclusion of DESI+$r_d$ systematically reduces uncertainties while preserving the SNIa-driven trends.

When it comes to the snap parameter, by contrast, it is most strongly constrained by BAO. This reflects the fact that $s(z)$ affects most directly the expansion rate $H(z)$ in our parameterization, and thus impacts BAO measurements of $D_H$ more strongly than it affects SNIa luminosity distances. SNIa constraints on $s(z)$ are therefore broader, while BAO consistently provide sharper measurements. However, despite the larger uncertainties of SNIa, we again find overall consistency between probes, with the combined datasets delivering robust constraints on $s(z)$ at the few-percent level.

The cosmographic reconstruction of $H(z)$ itself is shown both in Fig.~\ref{fig:S2} and in more detail in Fig.~\ref{fig:S3}. In the latter figure, the cosmographic points are compared with $\Lambda$CDM (blue bands) and CPL (green bands), which, as in the main text, are derived from Eqs.\eqref{eq:qz0_LCDM}-\eqref{eq:sz0_LCDM} and Eqs.\eqref{eq:qz_CPL}-\eqref{eq:jz_CPL}, respectively. The trend observed in the figure directly supports the conclusion drawn in the main text: at $z\gtrsim0.2$, the reconstructed expansion rate lies systematically above the $\Lambda$CDM prediction by about $2$ km/s/Mpc. The residuals in the bottom panel make this shift evident: all dataset combinations consistently prefer $H(z)$ values larger than $\Lambda$CDM, and the alignment with CPL predictions is visibly closer. This behavior is stable across pivots and is seen both in DESI+$r_d$+PP+CC and DESI+$r_d$+DESy5+CC, confirming that it is not driven by a single dataset but represents a coherent feature across independent probes.

Taken together, these supplementary results strengthen the conclusions of the main text. Not only do the different probes provide mutually consistent constraints, but their complementarity allows us to sharpen the reconstruction of late-time cosmic kinematics: BAO and SNIa both contribute significantly to $q(z_0)$ and $H(z_0)$; SNIa dominate the sensitivity to the jerk parameter at intermediate redshifts; BAO provide the sharpest constraints on the snap parameter; and the combined analyses consistently improve precision while preserving coherence. The overall consistency across probes, combined with the coherent redshift-dependent trends of the cosmographic parameters, provides strong evidence for the robustness of the framework and indicates that the observed deviations can hardly be attributed to noise or to any single dataset.

\section{Implications for the Hubble tension}
\label{section:D}

A key aspect of our cosmographic reconstruction is that it directly traces the redshift evolution of the Hubble parameter without assuming any specific cosmological model. This allows us to examine, in a fully model-independent manner, how the late-time expansion history inferred from DESI BAO, SNIa and CC relates to the long-standing Hubble tension.

The extrapolated value of the expansion rate at $z=0$ obtained from the Padé series is $H_0 = 67.49 \pm 0.56$ km/s/Mpc for DESI+$r_d$+PP+CC, and $H_0 = 66.90 \pm 0.53$ km/s/Mpc for DESI+$r_d$+DESy5+CC. An immediate observation is that these values are systematically lower than the $H_0$ inferred from the same datasets under the assumption of $\Lambda$CDM by roughly 1 km/s/Mpc, while being fully consistent with the $H_0$ inferred assuming a dynamical dark energy model such as CPL. This is clearly visible in the bottom panels of Fig.~\ref{fig:S3}: the cosmographic residuals (red) closely match the CPL predictions (green), whereas the $\Lambda$CDM curve (blue) lies noticeably higher at $z=0$.

This behavior reinforces a trend already highlighted in the recent literature: although dynamical dark energy models are statistically favored over a cosmological constant, they do not alleviate the Hubble tension~\citep{DESI:2025zgx,Giare:2025pzu,Ozulker:2025ehg}. On the contrary, they typically exacerbate it by shifting the inferred late-time $H_0$ to even lower values. This naturally raises the question of how models proposed to resolve the tension might affect our cosmographic analysis and the conclusions drawn from it.

To answer this question, we note that although our approach is model-independent, it is still sensitive to how the observational data are calibrated, most notably through the sound horizon used to anchor BAO and the absolute magnitude calibration used for SNIa. Any mechanism able to address the Hubble tension can be broadly classified into two categories, which affect these calibrations differently:

\begin{itemize}
\item \textbf{Early-time new physics:} These scenarios modify the expansion history before recombination, reducing the sound horizon $r_d$ and increasing the inferred value of $H_0$. Early Dark Energy (EDE) is a well-known example, typically requiring a 5–7\% reduction in $r_d$ to shift $H_0$ toward the SH0ES value, see, e.g., ~\citet{Poulin:2023lkg}.

\item \textbf{Late-time new physics:} These scenarios modify the expansion history after recombination, leaving the sound horizon essentially unchanged but increasing $H_0$ at low redshift. In these models, $r_d$ usually remains consistent with $\Lambda$CDM, but the absolute expansion rate today shifts the inferred $H_0$ toward the SH0ES value.
\end{itemize}

These two classes of solutions have distinct implications for our cosmographic framework. Early-time solutions require modifying the BAO calibration, whereas late-time solutions can be represented by recalibrating SNIa distances using the SH0ES determination of $H_0$, thus remaining agnostic on $r_d$. To explore these effects in a model-independent way, without committing to any specific new-physics scenario, we adopt two proxy implementations:

\begin{itemize}
\item \textbf{Proxy for early-time solutions:} we repeat the full cosmographic analysis using an EDE-like prior on the sound horizon $r_d^{\rm SH0ES}=139.00\pm 0.26$ Mpc, that is representative of the reduction required by early-time solutions to significantly alleviate the Hubble tension. We therefore consider the combinations DESI+PP+$r_d^{\rm SH0ES}$ and DESI+DESy5+$r_d^{\rm SH0ES}$, which test the sensitivity of our conclusions to changes in the BAO anchor.

\item \textbf{Proxy for late-time solutions:} we impose the SH0ES $H_0$ (or equivalently $M_b$) prior to calibrate the SNIa absolute magnitude~\citep{Riess:2021jrx} while leaving BAO uncalibrated (i.e.\ without any prior on $r_d$). We therefore consider the combinations DESI+PP+SH0ES and DESI+DESy5+SH0ES, which probe how the SH0ES calibration of SNIa affects the results of our cosmographic analysis.
\end{itemize}

\begin{table*}[htpb!]
\centering
\caption{Constraints at 68\% CL on the expansion rate $H(z_0)$ (in km/s/Mpc) and on the cosmographic parameters $q(z_0)$, $j(z_0)$, and $s(z_0)$ at different fixed pivot redshifts $z_0$, for various combinations of datasets. The table compares the impact of two distinct calibrations: \textit{(i)} a reduced sound horizon $r_d^{\rm SH0ES}=139\pm0.26$ Mpc, representative of early-time solutions to the Hubble tension, and \textit{(ii)} the SH0ES calibration of SNIa. For PP+SH0ES we directly adopt the SH0ES SNIa calibration, while for DESy5+SH0ES we impose an external Gaussian prior on $H_0$ centred on the SH0ES value of $73.04\pm1.04$ km/s/Mpc.}
\vspace{0.1cm}
\renewcommand{\arraystretch}{0.7}
\resizebox{0.9\textwidth}{!}{
\begin{tabular}{l @{\hspace{0.4 cm}} l @{\hspace{0.4 cm}}c@{\hspace{0.4 cm}}c@{\hspace{0.4 cm}}c@{\hspace{0.4 cm}}c}
\hline\hline
\\[-1ex]
\textbf{Redshift} & \textbf{Dataset} & \boldmath{$H(z_0)$} & \boldmath{$q(z_0)$} & \boldmath{$j(z_0)$} & \boldmath{$s(z_0)$} \\
\\[-1ex]
\hline
&&&&&\\
$z_0=0.0$ & DESI+PP+$r_{d_\text{SH0ES}}$ & $71.39\pm 0.61$ & $-0.423^{+0.058}_{-0.051}$ & $0.75^{+0.23}_{-0.30}$ & $-0.28^{+0.22}_{-0.48}$ \\
           & DESI+DESY5+$r_{d_\text{SH0ES}}$ & $70.81\pm 0.55$ & $-0.367^{+0.046}_{-0.041}$ & $0.55^{+0.16}_{-0.21}$ & $-0.47^{+0.10}_{-0.16}$ \\
           & DESI+PP+SH0ES & $73.5\pm 1.0$ & $-0.441^{+0.068}_{-0.056}$ & $0.85^{+0.27}_{-0.38}$ & $-0.11^{+0.35}_{-0.81}$ \\
           & DESI+DESY5+SH0ES & $73.0\pm 1.1$ & $-0.365^{+0.045}_{-0.040}$ & $0.54^{+0.15}_{-0.19}$ & $-0.488^{+0.076}_{-0.12}$ \\[1ex]
\hline\\[-1ex]
$z_0=0.1$ & DESI+PP+$r_{d_\text{SH0ES}}$ & $75.78\pm 0.49$ & $-0.377\pm 0.038$ & $0.86^{+0.22}_{-0.26}$ & $-0.45^{+0.25}_{-0.45}$ \\
           & DESI+DESY5+$r_{d_\text{SH0ES}}$ & $75.47\pm 0.46$ & $-0.323\pm 0.034$ & $0.62^{+0.17}_{-0.20}$ & $-0.673^{+0.093}_{-0.15}$ \\
           & DESI+PP+SH0ES & $77.7\pm 1.1$ & $-0.385\pm 0.040$ & $0.91^{+0.24}_{-0.27}$ & $-0.37^{+0.31}_{-0.52}$ \\
           & DESI+DESY5+SH0ES & $77.6\pm 1.1$ & $-0.324\pm 0.035$ & $0.62^{+0.17}_{-0.21}$ & $-0.66^{+0.12}_{-0.18}$ \\[1ex]
\hline\\[-1ex]
$z_0=0.2$ & DESI+PP+$r_{d_\text{SH0ES}}$ & $80.27\pm 0.45$ & $-0.322\pm 0.032$ & $0.98^{+0.24}_{-0.27}$ & $-0.53^{+0.38}_{-0.62}$ \\
           & DESI+DESY5+$r_{d_\text{SH0ES}}$ & $80.24\pm 0.43$ & $-0.274\pm 0.029$ & $0.69^{+0.18}_{-0.21}$ & $-0.87^{+0.12}_{-0.20}$ \\
           & DESI+PP+SH0ES & $82.3\pm 1.2$ & $-0.313\pm 0.031$ & $0.89\pm 0.22$ & $-0.70^{+0.24}_{-0.41}$ \\
           & DESI+DESY5+SH0ES & $82.3\pm 1.4$ & $-0.274\pm 0.027$ & $0.69^{+0.17}_{-0.20}$ & $-0.88^{+0.11}_{-0.17}$ \\[1ex]
\hline\\[-1ex]
$z_0=0.3$ & DESI+PP+$r_{d_\text{SH0ES}}$ & $85.41\pm 0.40$ & $-0.236\pm 0.023$ & $0.72\pm 0.12$ & $-1.178^{+0.081}_{-0.11}$ \\
           & DESI+DESY5+$r_{d_\text{SH0ES}}$ & $85.06\pm 0.44$ & $-0.236\pm 0.027$ & $0.90\pm 0.23$ & $-0.93^{+0.28}_{-0.48}$ \\
           & DESI+PP+SH0ES & $87.8\pm 1.3$ & $-0.229\pm 0.026$ & $0.67^{+0.14}_{-0.16}$ & $-1.176^{+0.080}_{-0.11}$ \\
           & DESI+DESY5+SH0ES & $86.2\pm 1.7$ & $-0.236\pm 0.027$ & $0.90\pm 0.23$ & $-0.94^{+0.27}_{-0.45}$ \\[1ex]
\hline\\[-1ex]
$z_0=0.4$ & DESI+PP+$r_{d_\text{SH0ES}}$ & $90.83\pm 0.34$ & $-0.164\pm 0.018$ & $0.591\pm 0.043$ & $-1.446\pm 0.067$ \\
           & DESI+DESY5+$r_{d_\text{SH0ES}}$ & $90.65\pm 0.36$ & $-0.167\pm 0.021$ & $0.696\pm 0.091$ & $-1.459^{+0.091}_{-0.11}$ \\
           & DESI+PP+SH0ES & $93.6\pm 1.3$ & $-0.151\pm 0.019$ & $0.497\pm 0.069$ & $-1.349^{+0.094}_{-0.11}$ \\
           & DESI+DESY5+SH0ES & $92.5\pm 1.6$ & $-0.166\pm 0.020$ & $0.694\pm 0.090$ & $-1.467^{+0.089}_{-0.099}$ \\[1ex]
\hline\\[-1ex]
$z_0=0.5$ & DESI+PP+$r_{d_\text{SH0ES}}$ & $96.38\pm 0.33$ & $-0.112\pm 0.018$ & $0.607\pm 0.026$ & $-1.703\pm 0.079$ \\
           & DESI+DESY5+$r_{d_\text{SH0ES}}$ & $96.31\pm 0.34$ & $-0.108\pm 0.019$ & $0.653\pm 0.037$ & $-1.724^{+0.094}_{-0.11}$ \\
           & DESI+PP+SH0ES & $99.1\pm 1.4$ & $-0.103\pm 0.018$ & $0.570\pm 0.034$ & $-1.665\pm 0.078$ \\
           & DESI+DESY5+SH0ES & $99.0\pm 1.6$ & $-0.108\pm 0.019$ & $0.653\pm 0.039$ & $-1.722^{+0.095}_{-0.11}$ \\[1ex]
\hline\\[-1ex]
$z_0=0.6$ & DESI+PP+$r_{d_\text{SH0ES}}$ & $102.15\pm 0.35$ & $-0.062\pm 0.019$ & $0.652\pm 0.025$ & $-1.99^{+0.11}_{-0.13}$ \\
           & DESI+DESY5+$r_{d_\text{SH0ES}}$ & $102.11\pm 0.36$ & $-0.058\pm 0.019$ & $0.680\pm 0.026$ & $-1.99^{+0.13}_{-0.15}$ \\
           & DESI+PP+SH0ES & $104.8\pm 1.6$ & $-0.055\pm 0.019$ & $0.636\pm 0.027$ & $-1.95^{+0.11}_{-0.14}$ \\
           & DESI+DESY5+SH0ES & $105.4\pm 1.7$ & $-0.056\pm 0.019$ & $0.681\pm 0.027$ & $-1.99^{+0.13}_{-0.16}$ \\[1ex]
\hline\\[-1ex]
$z_0=0.7$ & DESI+PP+$r_{d_\text{SH0ES}}$ & $108.24\pm 0.42$ & $-0.010\pm 0.019$ & $0.709\pm 0.027$ & $-2.26^{+0.17}_{-0.22}$ \\
           & DESI+DESY5+$r_{d_\text{SH0ES}}$ & $108.24\pm 0.40$ & $-0.007\pm 0.019$ & $0.729\pm 0.027$ & $-2.26^{+0.18}_{-0.23}$ \\
           & DESI+PP+SH0ES & $110.9\pm 1.8$ & $-0.005\pm 0.019$ & $0.705\pm 0.029$ & $-2.23^{+0.17}_{-0.22}$ \\
           & DESI+DESY5+SH0ES & $112.0\pm 1.9$ & $-0.006\pm 0.021$ & $0.731^{+0.028}_{-0.032}$ & $-2.22^{+0.23}_{-0.31}$ \\[1ex]
\hline\\[-1ex]
$z_0=0.8$ & DESI+PP+$r_{d_\text{SH0ES}}$ & $114.59\pm 0.50$ & $0.040\pm 0.021$ & $0.773\pm 0.033$ & $-2.51^{+0.28}_{-0.37}$ \\
           & DESI+DESY5+$r_{d_\text{SH0ES}}$ & $114.64\pm 0.47$ & $0.045\pm 0.019$ & $0.791\pm 0.030$ & $-2.46^{+0.30}_{-0.41}$ \\
           & DESI+PP+SH0ES & $117.3\pm 1.9$ & $0.046\pm 0.021$ & $0.776\pm 0.034$ & $-2.43^{+0.31}_{-0.44}$ \\
           & DESI+DESY5+SH0ES & $118.6\pm 2.3$ & $0.046\pm 0.021$ & $0.793\pm 0.032$ & $-2.43^{+0.34}_{-0.46}$ \\[1ex]
\hline\\[-1ex]
$z_0=0.9$ & DESI+PP+$r_{d_\text{SH0ES}}$ & $121.38\pm 0.63$ & $0.095\pm 0.022$ & $0.850\pm 0.039$ & $-2.59^{+0.52}_{-0.73}$ \\
           & DESI+DESY5+$r_{d_\text{SH0ES}}$ & $121.33\pm 0.64$ & $0.094\pm 0.024$ & $0.858\pm 0.040$ & $-2.58^{+0.55}_{-0.79}$ \\
           & DESI+PP+SH0ES & $124.0\pm 2.1$ & $0.099\pm 0.023$ & $0.855\pm 0.040$ & $-2.44^{+0.59}_{-0.84}$ \\
           & DESI+DESY5+SH0ES & $124.7^{+4.0}_{-3.0}$ & $0.098\pm 0.021$ & $0.865\pm 0.036$ & $-2.51^{+0.54}_{-0.74}$ \\[1ex]
\hline\\[-1ex]
$z_0=1.0$ & DESI+PP+$r_{d_\text{SH0ES}}$ & $128.1^{+1.4}_{-0.96}$ & $0.134^{+0.047}_{-0.033}$ & $0.910^{+0.082}_{-0.059}$ & $-2.73^{+0.96}_{-1.3}$ \\
           & DESI+DESY5+$r_{d_\text{SH0ES}}$ & $128.51\pm 0.69$ & $0.151\pm 0.023$ & $0.947\pm 0.043$ & $-2.38^{+0.86}_{-1.0}$ \\
           & DESI+PP+SH0ES & $130.8\pm 2.1$ & $0.151\pm 0.023$ & $0.938\pm 0.042$ & $-2.39^{+0.84}_{-1.0}$ \\
           & DESI+DESY5+SH0ES & $128.4^{+9.4}_{-4.0}$ & $0.150\pm 0.022$ & $0.946\pm 0.040$ & $-2.43^{+0.85}_{-0.97}$ \\[1ex]
\hline
\bottomrule
\end{tabular}}
\label{tab:results_SH0ES}
\end{table*}

\begin{figure}[htpb!]
    \centering
    \includegraphics[width=0.9\linewidth]{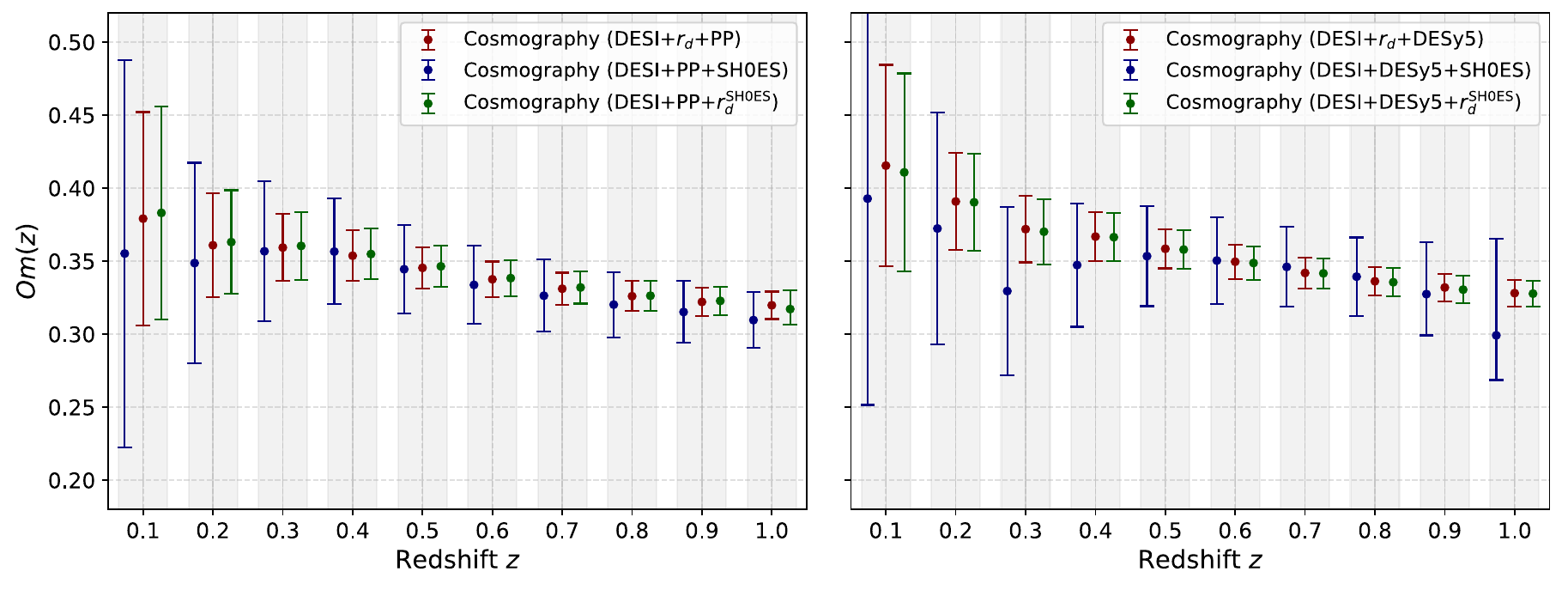}
    \includegraphics[width=0.9\linewidth]{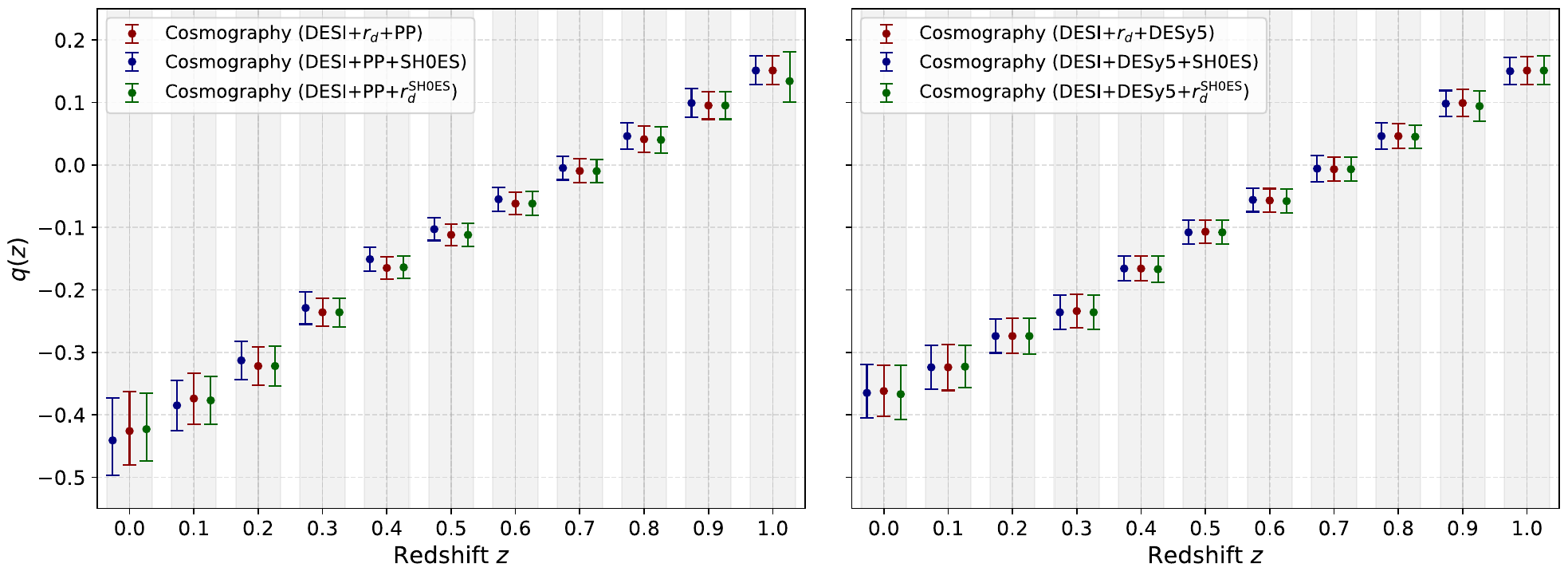}
    \includegraphics[width=0.9\linewidth]{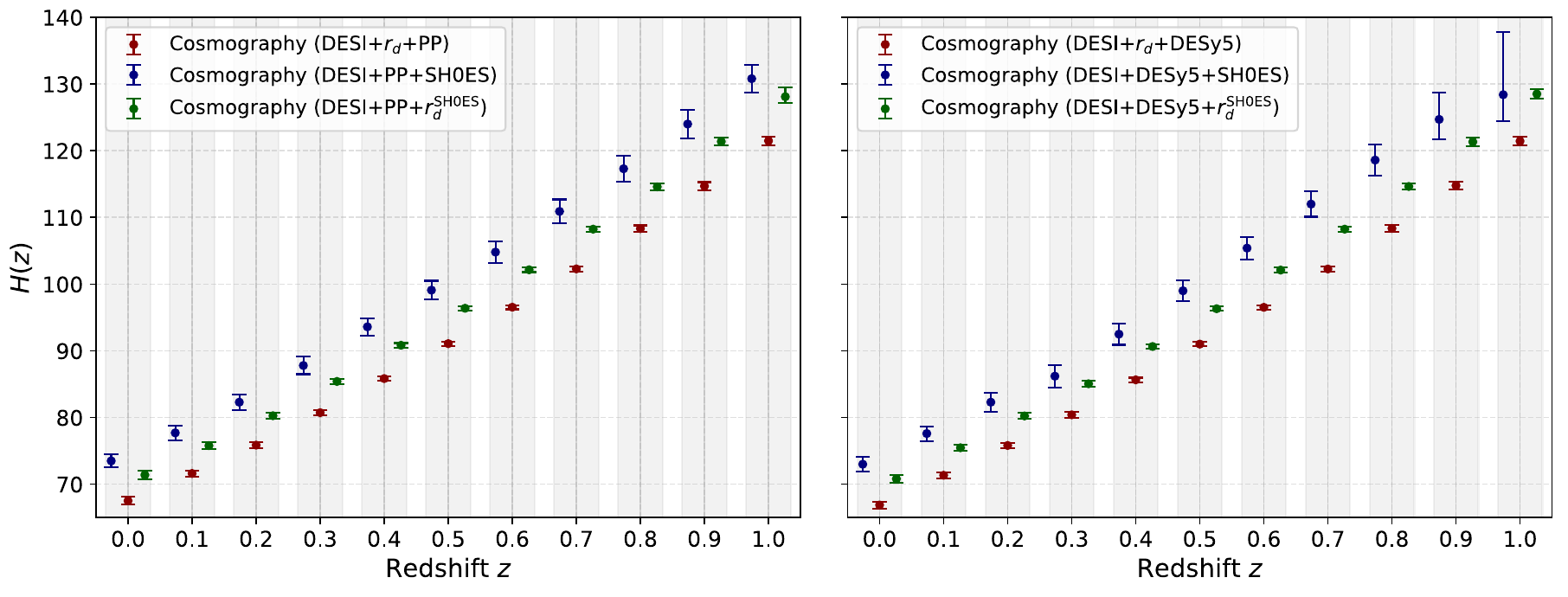}
    \caption{Whisker plot summarizing the 68\% CL constraints on the cosmographic parameters $Om(z)$, $q(z)$ and $H(z_0)$ obtained by \textit{(i)} calibrating BAO with the Planck sound horizon (red points), \textit{(ii)} calibrating SNIa with the SH0ES determination of $H_0$ while leaving BAO uncalibrated (blue points), and \textit{(iii)} calibrating BAO with the reduced sound horizon $r_d^{\rm SH0ES}=139\pm0.26$ Mpc representative of early-time solutions (green points). As discussed in the text, these three calibrations induce an overall shift in the amplitude of $H(z)$ but leave the shape-dependent quantities $q(z)$ and $Om(z)$ unchanged, leading to mutually consistent constraints across all pivot redshifts.}
    \label{fig:S4}
\end{figure}

These two procedures allow us to map, in a controlled and model-agnostic way, how our cosmographic reconstruction and the inferred preference for dynamical dark energy respond to the types of modifications typically invoked to address the Hubble tension. The numerical results obtained for all cosmographic parameters in the SH0ES-calibrated cases are reported in Tab.~\ref{tab:results_SH0ES} and visualized in Fig.~\ref{fig:S4}. In the figure, the red points in the left (right) panels correspond to DESI+$r_d$+PP (DESI+$r_d$+DESy5), i.e.\ BAO calibrated with the Planck $r_d$ value assuming $\Lambda$CDM. The blue points correspond to DESI+PP+SH0ES (DESI+DESy5+SH0ES), where SNIa are calibrated with the SH0ES determination of $H_0$ (or $M_b$) and no prior is imposed on $r_d$. Finally, the green points correspond to DESI+$r_d^{\rm SH0ES}$+PP (DESI+$r_d^{\rm SH0ES}$+DESy5), where BAO are calibrated with the reduced sound horizon $r_d^{\rm SH0ES}=139\pm 0.26$ Mpc discussed above.

As shown in the bottom panel of Fig.~\ref{fig:S4} (and in the corresponding $H(z_0)$ values in Tab.~\ref{tab:results_SH0ES}), adopting either an SH0ES-based calibration for SNIa or an EDE-like prior on $r_d^{\rm SH0ES}$ shifts all reconstructed values of $H(z)$ upward. As expected, the SH0ES-calibrated SNIa produce the highest values of $H_0$, fully consistent with SH0ES, while the $r_d^{\rm SH0ES}$ prior yields intermediate values that move significantly closer to SH0ES. We also note that the prior on $r_d^{\rm SH0ES}$ has a much smaller uncertainty than the SH0ES prior on $H_0$, which explains why the green points display smaller error bars than the blue ones. Overall, the net effect on the expansion rate can be summarized as an overall rescaling,
\begin{equation}
H(z)^{\rm SH0ES} = \alpha\, H(z)^{\rm Planck},
\end{equation}
between the values inferred from cosmography using either of the SH0ES-based calibrations (green and blue points) and those inferred when BAO are calibrated with the Planck sound horizon (red points). This behavior is clearly visible in the bottom panel of Fig.~\ref{fig:S4}.

We now show, both explicitly and analytically, that such a rescaling has no impact on the two key quantities used in the main text to assess the preference for dynamical dark energy: the deceleration parameter $q(z)$ and the $Om(z)$ diagnostic. This is immediately evident from the middle and top panels of Fig.~\ref{fig:S4}. The constraints on $q(z)$ and $Om(z)$ obtained from SH0ES-calibrated data (blue and green), and Planck-calibrated data (red) remain in excellent mutual agreement across all redshift bins. This confirms that the upward shift in $H(z)$ does not affect the quantities most relevant to assessing deviations from $\Lambda$CDM.

The reason is that both $q(z)$ and $Om(z)$ depend only on the \emph{shape} of $H(z)$, not on its overall amplitude. In more quantitative terms, $H_0$ sets the normalization of $H(z)$, but by definition
\begin{equation}
q(z) = \frac{d \ln H}{d \ln (1+z)} - 1 = \frac{1}{H(z)} \frac{d H}{d \ln (1+z)} - 1,
\end{equation}
is invariant under the rescaling $H(z) \rightarrow \alpha\, H(z)$ as one can write
\begin{equation}
\frac{1}{H(z)^{\rm SH0ES}} 
\frac{d H(z)^{\rm SH0ES}}{d \ln (1+z)}
= 
\frac{1}{\alpha H(z)^{\rm Planck}} 
\frac{d (\alpha H(z)^{\rm Planck})}{d \ln (1+z)}
=
\frac{1}{H(z)^{\rm Planck}} 
\frac{d H(z)^{\rm Planck}}{d \ln (1+z)},
\end{equation}
which shows explicitly that $q(z)$ is unchanged by such a shift. Similarly, the $Om(z)$ diagnostic is defined as
\begin{equation}
Om(z) \equiv \frac{\tilde{h}^2(z)-1}{(1+z)^3-1},
\end{equation}
with $\tilde{h}(z)=H(z)/H_0$. Under a SH0ES-like rescaling one has
\begin{equation}
\tilde{h}(z)^{\rm SH0ES}
= \frac{H(z)^{\rm SH0ES}}{H_0^{\rm SH0ES}}
= \frac{\alpha H(z)^{\rm Planck}}{\alpha H_0^{\rm Planck}}
= \tilde{h}(z)^{\rm Planck},
\end{equation}
and therefore $Om(z)$ is also invariant.

Altogether, this exercise shows that the cosmographic results for $q(z)$ and $Om(z)$ are essentially unaffected by new physics capable of shifting the absolute expansion scale, either through an early-time reduction of the sound horizon (as in EDE-like scenarios) or through a late-time recalibration consistent with SH0ES. As a result, the kinematic evidence for departures from $\Lambda$CDM reported in the main text is not tied to the particular calibration adopted, and remains fully robust under the types of modifications typically invoked to address the Hubble tension.

\newpage
\bibliographystyle{aasjournalv7}
\bibliography{main}

\end{document}